# Giant exciton effects and magneto-excitonic coupling in $V_4S_9X_4$ 2D magnetic semiconductors

Yingjie Wei[1], Fan Zhang[2], Ying Zhao[3], Lixin Zhou[1], Yan Su[1], Yu Guo[1*], Jijun Zhao[4,5*]

[1] Key Laboratory of Materials Modification by Laser, Ion and Electron Beams (Dalian University of Technology), Ministry of Education, Dalian 116024, China

[2] School of Physics, Nanjing University, Nanjing 210093, China

[3] School of Electronic Engineering, Huainan Normal University, 232038, China

[4] Guangdong Provincial Key Laboratory of Quantum Engineering and Quantum Materials, School of Physics, South China Normal University, Guangzhou 510006, China

[5] Guangdong Basic Research Center of Excellence for Structure and Fundamental Interactions of Matter, South China Normal University, Guangzhou 510006, China

[*]Corresponding authors. Email: guoyu_dlut@dlut.edu.cn, zhaojj@scnu.edu.cn

## Abstract

Room-temperature spin-optoelectronic devices require a combination of robust ferromagnetism and giant exciton binding—a pairing mutually exclusive in conventional semiconductors due to magnetic localization that screens excitons. Cluster-assembled $V_4S_9X_4$ (X = F, Cl, Br and I) monolayers overcome this bottleneck via a hierarchical design, that is, intra-cluster localized states host both local magnetic moments and strong electron-hole interactions, while inter-cluster coupling mediates long-range ferromagnetism. Remarkably, these two-dimensional semiconductors exhibit intrinsic ferromagnetism with Curie temperature up to 507.6 K. As a prototype, $V_4S_9Br_4$ monolayer possesses a giant exciton binding energy of 1.85 eV. Its lowest exciton is a dark state ($D_I$) with a radiative lifetime of 1.20 ns, whereas the first bright exciton ($B_I$) exhibits an ultrafast radiative decay of 86.87 ps. This stark lifetime contrast enables simultaneous ultrafast optical response and long-lived spin information storage. Most notably, switching between ferromagnetic and antiferromagnetic order allows for wide-range tuning of exciton lifetime, with the giant binding energy remaining nearly intact. Our findings establish cluster assembly as a powerful paradigm for designing next-generation spin-photonic and quantum information devices operating at room temperature.

## Introduction

Seamlessly merging intrinsic room-temperature ferromagnetism with robust exciton binding is paramount for advancing spin-optoelectronic technologies. This synergy unlocks zero-field magnetic manipulation of exciton spin dynamics while ensuring thermally stable excitonic operations, thereby eliminating the need for cryogenic cooling and intense external magnetic fields mandated by conventional schemes.[1] Yet, reconciling these two properties poses a formidable physical conflict: robust magnetic order dictates highly localized electronic states, while giant exciton binding inherently requires weak dielectric screening. As a result, current material platforms are restricted to a trade-off. Mainstream two-dimensional (2D) semiconductors host strongly bound excitons but remain mostly nonmagnetic,[2, 3] whereas 2D ferromagnets typically exhibit fragile exciton binding and lackluster room-temperature optical activity.[4, 5]

Beyond the mere coexistence of ferromagnetism and strongly bound excitons, a more fundamental question is how specific magnetic orders—ferromagnetic (FM) versus antiferromagnetic (AFM)—modulate the exciton properties. Early investigations on 2D magnetic semiconductors, such as $NiBr_2$,[6] CrSBr[7, 8] and $MnPS_3$,[9] have revealed an intimate entanglement between magnetic configurations and excitonic profiles. For instance, switching between FM and AFM configurations alters the exciton binding energy by hundreds of meV in $NiBr_2$ and $MnPS_3$. Meanwhile, the onset of AFM order in CrSBr confines excitons to individual layers, resulting in a sharp narrowing of their linewidths.[10] This sensitivity underscores the intense coupling between spin alignment and electron-hole interactions in conventional magnetic systems. However, it remains an open question whether orthogonal control can be achieved—namely, maintaining a robust and giant exciton binding energy invariant across distinct magnetic phases, while simultaneously exploiting magnetic order to selectively modulate other key excitonic properties, such as lifetime and spatial localization.

To address this challenge, cluster assembly[11–14]—a bottom-up paradigm bridging molecular chemistry and solid-state physics—offers a powerful route to circumvent this long-standing limitation. By employing atomically precise transition metal chalcogenide clusters as building blocks, this strategy successfully decouples intra-cluster and inter-cluster interactions. Specifically, robust local magnetic moments and intense Coulomb interactions are engineered within individual clusters, whereas weak inter-cluster coupling sustains long-range magnetic order in a low-dielectric environment favorable for giant exciton binding. This hierarchical electronic architecture provides an ideal platform to decouple magnetic order from exciton binding energy—a capability intrinsically unattainable in conventional covalent materials. Vanadium thiobromide ($V_4S_9Br_4$) serves as a prototypical embodiment of this design, featuring halogen-bridged $[V_4S_9]^{4+}$ clusters assembled into a layered framework.[15] Yet, while its bulk form has been synthesized and magnetically characterized,[16, 17] its monolayer exfoliation and fundamental excitonic behavior remain entirely unexplored.

Herein we perform systematic first-principles calculations to investigate the family of cluster-assembled $V_4S_9X_4$ (X = F, Cl, Br and I) monolayers. We demonstrate that they are intrinsic ferromagnetic semiconductors with Curie temperatures well above or near room temperature. Taking $V_4S_9Br_4$ monolayer as a representative prototype, we unveil exceptionally strong excitonic effects, characterized by a giant exciton binding energy of 1.85 eV and a distinct energy hierarchy between dark and bright excitons. Crucially, a comparative analysis between FM and AFM configurations demonstrates that the magnetic order can effectively modulate exciton radiative lifetimes and spatial localization, all while leaving the robust binding strength virtually intact. Collectively, these findings establish halogen-bridged cluster architectures as a versatile and promising platform for engineering magneto-excitonic materials with orthogonally tunable functionalities.

## Computational methods

All quantum chemical calculations for the $[V_4S_9]^{4+}$ cluster, including geometry optimization and electronic properties evaluation, were carried out using the Gaussian 16 software suite[18] within the framework of density functional theory (DFT). The Perdew-Burke-Ernzerhof (PBE) functional[19] within generalized gradient approximation (GGA) was adopted in conjunction with the aug-cc-pvdz basis set.[20] To ensure rigorous computational accuracy, the energy convergence threshold was set to $10^{-6}$ Hartree. Furthermore, time-dependent density functional theory (TDDFT)[21] calculations were conducted to determine the excited-state energy levels and oscillator strengths of the $[V_4S_9]^{4+}$ cluster.

Spin-polarized DFT calculations for the 2D cluster-assembled monolayers were conducted using the Vienna *ab initio* Simulation Package (VASP)[22] with the PBE functional.[19] The projector augmented wave (PAW) approach[23] modeled the ion–electron interaction, utilizing a plane-wave kinetic energy cutoff of 500 eV. Structural relaxations proceeded until the energy and force tolerances reached $10^{-5}$ eV/atom and $10^{-3}$ eV/Å, respectively, under an interlayer vacuum spacing exceeding 24 Å to preclude spurious periodic interactions. A Monkhorst-Pack k-point mesh with a reciprocal resolution of $2\pi\times0.03$ Å$^{-1}$ sampled the Brillouin zone. To assess the dynamical stability, phonon dispersion spectra were computed using the finite displacement method implemented in PHONOPY.[24, 25] Given the relatively light mass of vanadium, spin-orbit coupling was neglected for electronic and excitonic structures calculations—as validated by our preliminary benchmarks (Fig. S1)—and was uniquely incorporated for magnetocrystalline anisotropy energy (MAE) calculations.

Quasiparticle band structures and excitonic effects were determined using the BerkeleyGW package[26–28] with mean-field electronic structure provided by the Quantum ESPRESSO software.[29] Quasiparticle energies were obtained at the one-shot $G_0W_0$ level, and subsequent Bethe–Salpeter equation (BSE) calculations were performed to resolve exciton states. Coulomb interactions were truncated in the out-of-plane direction to account for the 2D nature of monolayers. All details for these calculations are given in the Supplemental Materials (Table S1 and Fig. S2).

## Results and Discussion

### Structural design and stability of cluster-assembled monolayer

Given the successful synthesis of layered $V_4S_9Br_4$,[15] we herein systematically examine the structural framework and intrinsic stability of its isoelectronic counterparts (i.e., $V_4S_9X_4$) in monolayer form. The discrete $[V_4S_9]^{4+}$ polycation serves as the supramolecular building block of the system, featuring a $\{V_4\}$ core (Fig. 1). Within this central magnetic motif, four equivalent V ions possess a formal mixed-valence state of +3.5,[15] arranging into a regular planar square. Each $V^{3.5+}$ hosts an unpaired *d* electron that yields a local magnetic moment of 1 $\mu_B$, while the remaining *d* electrons occupy the localized V-V bonding orbitals. This $\{V_4\}$ core is encapsulated by three crystallographically distinct sulfur species, all exhibiting closed-shell $s^2p^6$ configurations: one central $\mu_4$-$S^{2-}$ ligand ($S_1$) coordinating all four V ions, and four edge-bridging $S_2^{2-}$ disulfide dimers ($S_2$, $S_3$) positioned on alternating sides of the square plane. Because all surrounding ligands are nonmagnetic, a total spin moment of ~4 $\mu_B$ for the cluster and the robust intra-cluster FM exchange stem exclusively from the partially filled V-*d* states. The complete absence of imaginary frequencies in its vibrational spectrum underscores the dynamical stability of this isolated cluster, validating its potential for bottom-up materials design.

The structural evolution from discrete $[V_4S_9]^{4+}$ clusters into an extended 2D $V_4S_9X_4$ framework complies with the closed-shell charge neutrality principle,[30, 31] mediated by negatively charged halide anions $X^-$ as bridging linkers. All halide ions adopt a closed-shell electronic configuration, thereby contributing nothing to the net magnetic moment. Mechanistically, each $[V_4S_9]^{4+}$ cluster exposes four peripheral coordination sites, each occupied by an $X^-$ ligand; in turn, every $X^-$ bridges two adjacent clusters in a corner-sharing topology. This 1:2 connectivity perfectly satisfies charge neutrality within the unit cell—comprising two $[V_4S_9]^{4+}$ clusters and eight $X^-$ ligands. The resulting continuous V-X-V linkages propagate the square network within the *ab* plane, serving as the pathways that mediate inter-cluster magnetic exchange. All optimized structures belong to the $D_{4h}$ point group symmetry, with optimized lattice parameters spanning from 10.14 to 11.34 Å (Table 1). Ultimately, this charge-directed assembly yields a series of isostructural $V_4S_9X_4$ monolayers that faithfully preserve the structural, electronic, and magnetic characteristics of their constituent clusters. The synthetic feasibility of these isostructural $V_4S_9X_4$ monolayers has been systematically assessed via four criteria: energetic, dynamic, and thermal stability, as well as mechanical exfoliation feasibility. Detailed calculation methods and full results of exfoliation energies are provided in Section S3 of the Supplemental Materials.

**Electronic structures and magnetic properties**

The unique electronic and magnetic properties of $V_4S_9X_4$ monolayers are governed directly by the preserved structural integrity of their $[V_4S_9]^{4+}$ building blocks. Electronic structure calculations reveal a characteristic flat-band feature near the Fermi level across all monolayer systems, originating predominantly from the localized V-3*d* orbitals of the $\{V_4\}$ cores, whereas contributions from the closed-shell $S^{2-}$, $S_2^{2-}$ and $X^-$ ligands are negligible. These $V_4S_9X_4$ monolayers are intrinsic semiconductors, yielding PBE band gaps of 0.71–0.76 eV (Fig. S6), which are refined to 1.06–1.41 eV using a more accurate HSE06 hybrid functional[32] (Table 1). The band gap widens monotonically with increasing halogen atomic number. This trend stems from the expansion of V-X bond length in the presence of heavier halogens, which diminishes *p*-*d* orbital overlap and weakens the hybridization between halide *p* and vanadium *d* states. Local density of states (LDOS) analysis (Fig. S6) further substantiates that the valence band maximum (VBM) comprises predominantly V-3*d* orbitals, while the conduction band minimum (CBM) is composed of hybridized V-3*d* and S-3*p* orbitals. Such pronounced orbital localization induces both large carrier effective masses (which suppress kinetic energy) and weak effective dielectric screening (which enhances Coulombic attraction). These factors collectively favor tightly bound excitons, laying the theoretical foundation for the giant exciton binding energy discussed later.

All $V_4S_9X_4$ monolayers favor FM ground state, as verified by energy comparisons between the FM and various AFM configurations (Fig. S7). Consistent with the intrinsic characteristics of the $[V_4S_9]^{4+}$ building block, the spin density is fully localized on the $V^{3.5+}$ centers, with negligible contributions from the closed-shell ligands. The local magnetic moment originates from the crystal-field splitting of V-3*d* orbitals under the pentagonal bipyramidal coordination environment at each V site. As illustrated in Fig. 2a, the fivefold-degenerate V-3$d$ orbitals split into three distinct energy levels within such ligand field: the highest-energy $a_1$' ($d_{z^2}$) orbital, the intermediate doubly degenerate $e_2$' ($d_{x^2-y^2}$, $d_{xy}$) orbitals, and the lowest-energy doubly degenerate $e_1$'' ($d_{yz}$, $d_{xz}$) orbitals. The single unpaired electron occupies the $e_1$'' orbital, yielding a net spin magnetic moment of ~1 $\mu_B$ per V atom. This highly localized electronic configuration underpins both the robust intra-cluster FM coupling and the large perpendicular magnetocrystalline anisotropy within the extended 2D lattice. Specifically, intra-cluster exchange aligns the spin moments at four V sites in parallel within each cluster, producing a total moment of ~4 $\mu_B$. Meanwhile, inter-cluster coupling, mediated by the bridging halide ligands, then stabilizes long-range FM order across the entire monolayer.

Magnetic exchange interactions are quantified via the four-state energy-mapping method, with total energies of four collinear magnetic configurations mapped onto a Heisenberg Hamiltonian:[33, 34]

$$H = -\sum_{i<j} J_{ij} M_i M_j \quad (1)$$

where $J_{ij}$ is the exchange coupling between sites $i$ and $j$, and $M_i$ is the local magnetic moment at site $i$. For all compositions, $J_1$ is positive and dominant (10.3–22.8 meV), corresponding to strong FM nearest-neighbor interactions. Longer-range couplings $J_2$, $J_3$, and $J_4$ are an order of magnitude weaker, with $J_3$ exhibiting weak AFM character. The physical origin of these exchange interactions is illustrated in Fig. 2c. Briefly, $J_1$ stems from two near-90° superexchange pathways ($V_1$-S-$V_2$ and $V_1$-X-$V_2$), in line with the Goodenough-Kanamori rules.[35–37] The monotonic reduction of $J_1$ from F to I is consistent with the lengthening V-X bond, as shorter bonds enhance $p$-$d$ orbital overlap, in turn strengthening superexchange interaction. This geometric control of the exchange coupling is further quantified by the systematic correlation between bond parameters and $J_1$ across the halogen series (Fig. S8). The antiferromagnetic sign of $J_3$ deviates from standard predictions due to the four-coordinate $\mu_4$-$S^{2-}$ bridging pathway. Direct intra-cluster V-V $d$-orbital exchange is negligible at equilibrium bond lengths (>2.8 Å), confirming that magnetism in these systems is entirely mediated by ligand superexchange.

The MAE was evaluated from the total energy difference between in-plane and out-of-plane magnetization directions. All systems exhibit an out-of-plane easy axis, yielding positive MAE values of 108–160 μeV/atom (Table 1). The Curie temperature ($T_C$) was estimated via Monte Carlo simulations with the extracted exchange parameters (Fig. 2b and Fig. S9).[38] Notably, the $T_C$ values closely mirror the trend of $J_1$: $V_4S_9F_4$ exhibits the highest $T_C$ of 507.6 K, followed by $V_4S_9Cl_4$ (276.5 K), $V_4S_9Br_4$ (268.3 K), and $V_4S_9I_4$ (275.0 K), with the latter three remaining near room temperature. These results establish that nearest-neighbor superexchange is the primary mechanism stabilizing the long-range FM order in these cluster-assembled systems.

**Quasiparticle electronic structure of $V_4S_9Br_4$ monolayer**

Motivated by the synthesis of layered $V_4S_9Br_4$,[15] we choose $V_4S_9Br_4$ monolayer as a representative to investigate the excitonic effects. Its quasiparticle band structure was computed using the GW approximation. As shown in Fig. 3a, ferromagnetic $V_4S_9Br_4$ exhibits a pronounced exchange splitting: the majority-spin (spin-up) and minority-spin (spin-down) channels feature bandgaps of 2.06 eV and 2.64 eV, respectively. Although the smaller majority-spin gap facilitates carrier population, its optical transitions are strongly constrained by orbital symmetry, as discussed below. Crucially, the bands near the Fermi level are exceptionally flat—a hallmark of the cluster-assembled lattice,[39] where weak inter-cluster electronic coupling localizes electronic states within individual $[V_4S_9]^{4+}$ units. This flat-band character yields heavy carrier effective masses: along X'–Γ–X, $m_h$ = 49.84 $m_0$ and $m_e$ = 22.99 $m_0$; along M'–Γ–M, $m_h$ = 46.63 $m_0$ and $m_e$ = 41.08 $m_0$ ($m_0$ is the free electron mass). These heavy masses play a central role in the giant exciton binding energy, which will be discussed later.

To elucidate the optical selection rules, we performed a group-theoretical analysis of the irreducible representations for valence and conduction bands at the direct-gap Γ point (Fig. 3b). Within $D_{4h}$ point group, the electric dipole operators transform as $A_{2u}$ ($z$-polarized) and $E_u$ ($x$, $y$-polarized). An optical transition is dipole-allowed only if the direct product of initial state, final state, and dipole operator representations contains the totally symmetric $A_{1g}$ representation. As summarized in Table 2 and Section S9 of Supplemental Materials, parity mismatch renders most low-energy interband transitions optically forbidden; specifically, transitions between states of the same parity (g→g or u→u) are dipole-forbidden, whereas those between opposite parities (g→u or u→g) are allowed.[40, 41] Consequently, the majority-spin channel permits only 3 out of 30 possible low-energy transitions, while the minority-spin channel allows 7. This symmetry-enforced dual-channel behavior provides a fundamental mechanism for the coexistence of dark and bright excitons in the $V_4S_9Br_4$ monolayer.

**Excitonic properties of $V_4S_9Br_4$ monolayer**

Consistent with symmetry analysis, the BSE-calculated absorption spectrum exhibits multiple discrete exciton peaks below the quasiparticle bandgap, with the lowest optically active peak corresponding to the first bright exciton ($B_I$). A comparison between the independent-particle approximation (IPA) and BSE absorption spectra (Fig. 4) highlights the dominant role of excitonic effects. While the IPA spectrum shows a smooth, featureless onset at the quasiparticle gap (2.06 eV), the BSE spectrum displays sharp discrete peaks well below the fundamental gap, redshifting the absorption edge by ~1.1 eV.[42] To trace the origin of the above features, we compared the monolayer spectrum with the TD-DFT spectrum of an isolated $[V_4S_9]^{4+}$ cluster (Fig. 4), which exhibits a weak exciton peak at ~0.65 eV. Upon assembling into a monolayer, analogous low-energy peaks emerge, albeit redshifted and split by bromide-mediated inter-cluster coupling. This resemblance confirms that excitons remain largely confined within individual $[V_4S_9]^{4+}$ units (Fig. 5).

Our BSE calculations yield a giant exciton binding energy ($E_B$) of 1.85 eV, defined as the energy difference between the majority-spin quasiparticle gap (2.06 eV) and the lowest exciton energy (0.21 eV). This ranks among the highest reported values for 2D semiconductors, far exceeding those of monolayer transition metal dichalcogenides (~0.5–0.6 eV).[43, 44] Such exceptionally large binding energies (>1.5 eV) for the lowest-lying excitons ensure robust excitonic effects at and above room temperature. Notably, the lowest-energy ground-state exciton ($D_I$) is dark, featuring a nearly vanishing optical transition dipole moment. This dark character is enforced by strict $D_{4h}$ parity selection rules and strong exchange coupling, aligning perfectly with our group-theoretical analysis. The first bright exciton ($B_I$) appears at ~0.31 eV with a substantial oscillator strength. A comparison of the excitonic spectra between the FM and metastable AFM phases is also provided in the following section to disentangle the influence of magnetic order.

**Microscopic origin of giant exciton binding energy**

The giant exciton binding energy in the $V_4S_9Br_4$ monolayer arises directly from its cluster-assembled hierarchical architecture, which simultaneously produces heavy carrier masses, weak dielectric screening, and strong spatial confinement. First, the ultra-flat band dispersion[45] yields a large exciton reduced mass, $\mu = (m_e \bullet m_h)/(m_e + m_h) = 15.7\ m_0$, which is nearly two orders of magnitude larger than that of conventional 2D semiconductors like monolayer $MoS_2$ (~0.19 $m_0$).[46] This heavy mass significantly quenches the kinetic energy of the electron-hole pair, thereby facilitating tighter Coulomb confinement. Concurrently, $V_4S_9Br_4$ monolayer exhibits a lower macroscopic electronic dielectric constant ($\varepsilon_\infty$ = 2.95; Table S3) compared to typical 2D systems ($WS_2$: ~14; $MoS_2$: ~15.4; $h$-BN: 4.97).[47–49] The weak dielectric screening further amplifies the effective Coulomb interaction between electrons and holes. Third, the excitonic wavefunctions are strongly confined within individual $[V_4S_9]^{4+}$ clusters. To characterize the exciton wavefunction, we expand it as a linear combination of independent electron-hole pair states within the Tamm-Dancoff approximation:[50]

$$\psi_S(r_e, r_h) = \sum_{c,v,k} A_{cvk}^S\, \psi_{ck}(r_e)\psi_{vk}^*(r_h) \quad (2)$$

where $S$ labels the exciton state, and $A_{cvk}^S$ is the electron-hole amplitude assigning weights to each transition from valence band ($v$) to conduction band ($c$) at momentum

$k$. The $k$-dependent spectral weight is defined as:

$$|\psi_S(k)|^2 = \sum_{c,v}\left|A_{cvk}^S\right|^2 = |A_S(k)|^2 \tag{3}$$

Wavefunction analysis in Fig. 5a confirms this strong confinement in the FM ground state, that is, excitons are predominantly localized within individual $[V_4S_9]^{4+}$ clusters, with finite inter-cluster delocalization visible in real space. The corresponding symmetric pattern in reciprocal space further indicates a mixed character, placing the excitons in an intermediate regime between the Wannier-Mott and Frenkel limits.[51, 52] Such intra-cluster confinement directly enhances the Coulomb coupling between charge carriers, thereby maximizing the attractive interaction that underlies the giant binding energy.[40]

**Dark-bright exciton splitting and exciton dynamics**

Building on the strong excitonic localization and giant binding energy, we evaluate the radiative lifetimes of dark ($D_I$) and bright ($B_I$) excitons using the transition dipole moments from BSE calculations. The radiative lifetime at 0 K is calculated as:[53, 54]

$$\tau = \frac{A_C h^2 c}{8\pi e^2 E_S \mu_S^2} \tag{4}$$

where $A_c$ denotes the unit cell area, $c$ represents the speed of light, $E_S$ is the exciton energy, and $\mu_S$ is the exciton transition dipole moment. Our calculations reveal a pronounced lifetime difference between dark and bright excitons. The ground-state dark exciton $D_I$ exhibits an exceptionally long radiative lifetime of 1.20 ns, which is a direct consequence of its vanishingly small transition dipole moment enforced by the strict parity selection rules of $D_{4h}$ point group. In sharp contrast, the first symmetry-allowed bright exciton $B_I$ possesses a much larger transition dipole moment, yielding an ultrashort radiative lifetime of 86.87 ps, which aligns perfectly with its prominent absorption intensity in the optical spectrum.

This intrinsic lifetime disparity between the long-lived dark excitons and ultrafast bright excitons is highly advantageous for spin-optoelectronic applications.[55] Specifically, the nanosecond-scale dark exciton $D_I$ provides an ample time window for spin manipulation and optical buffering, whereas the picosecond radiative decay of $B_I$ enables high-speed optical readout and switching. These complementary characteristics position $V_4S_9X_4$ monolayers as a promising platform for spin-photonic and quantum information devices. In the AFM metastable phase, the radiative lifetimes of both $D_I$ and $B_I$ are further prolonged. More broadly, as summarized in Table 3 and Fig. S10, dark excitons consistently exhibit radiative lifetimes orders of magnitude longer than their bright counterparts across both FM and AFM phases, which will be discussed in the next section.

**Intrinsic correlation between magnetism and excitons**

Having fully characterized the FM ground state of $V_4S_9Br_4$ monolayer, we now examine its AFM phase to isolate the impact of magnetic ordering. The AFM configuration exhibits a spin-degenerate indirect GW gap of 2.19 eV, merely 0.13 eV larger than the FM majority-spin direct gap (2.06 eV). The lowest AFM exciton resides at 0.41 eV, yielding a binding energy of 1.78 eV—only 0.07 eV lower than the FM value (1.85 eV). This insensitivity fundamentally contrasts with conventional 2D magnets (e.g., $NiBr_2$ and $MnPS_3$),[6, 9] underscoring the dominance of intra-cluster Coulomb interactions over long-range spin ordering.

Absorption spectra (Fig. 4b–c) and eigenstate analysis (Fig. 4d) show that magnetic order selectively reshapes low-energy excitons. In the FM monolayer, the low-energy feature redshifts to 0.2–0.3 eV and splits into dark ($D_I$, $D_{II}$) and bright ($B_I$) states, with the first bright exciton $B_I$ at 0.31 eV ($|\mu|^2 = 5.74$); the intra-cluster peak (~0.97 eV) remains nearly unchanged. In the AFM phase, weakly allowed states cluster at 0.41–0.44 eV, followed by a dark window (0.44–0.57 eV) with dipole moments suppressed by four orders of magnitude, and its first bright exciton emerges at ~0.67 eV, closely matching the isolated $[V_4S_9]^{4+}$ cluster. This ~0.36 eV blueshift of the bright exciton upon FM-to-AFM switching serves as a direct spectroscopic signature of strong magneto-exciton coupling.

Spatial wavefunction analysis (Fig. 5) further confirms the magnetic control of exciton character. The parallel spin alignment in the FM phase enables finite inter-cluster delocalization, yielding a mixed Frenkel-Wannier character.[56] Conversely, the alternating spins in the AFM phase block spin-conserving inter-cluster hopping, strictly confining excitons to localized cluster groups in a pure Frenkel fashion. This trend is corroborated by the reciprocal-space $|A_S(k)|^2$ distributions: the FM phase features a structured four-lobe profile—the Fourier signature of finite inter-cluster delocalization—whereas the AFM phase shows a nearly isotropic, diffuse pattern reflecting extreme real-space localization.

Despite similar binding energies, the FM-to-AFM transition drastically alters exciton dynamics. The $D_I$ and $B_I$ lifetimes elongate from 1.20 ns and 86.87 ps in the FM phase to 2.29 ns and 0.18 ns in the AFM phase, respectively, with the darkest parity-forbidden AFM exciton reaching an ultra-long 23.95 μs. Overall, AFM lifetimes exceed FM ones, and dark excitons outlive bright ones by orders of magnitude (Table 3, Fig. S10)—a trend driven by tighter AFM exciton confinement that enhances parity forbiddenness and suppresses radiative decay.[57, 58] Consistently, oscillator strength distributions (Fig. 4) reveal strong low-energy $B_I$ absorption in the FM phase but much weaker absorption in the AFM phase.

Magnetic order thus independently tunes exciton optical activity and lifetime without altering the giant binding energy, breaking the traditional trade-off in magneto-excitonic materials. Such unique tunability stems directly from the cluster-assembly paradigm. Unlike conventional 2D magnets—where strong magnetic exchange via orbital overlap enhances dielectric screening and delocalizes carriers, thereby suppressing binding energies—the cluster-assembled architecture circumvents this limitation through a hierarchical, two-scale design (Fig. 6). Here, individual clusters provide a localized electronic environment (large effective masses and weak screening) to sustain giant exciton binding. Simultaneously, inter-cluster bridging enables strong superexchange without compromising intra-cluster localization. Such decoupling of intra- and inter-cluster interactions effectively resolves the conflict between magnetism and excitonic effects. Ultimately, this strategy allows external fields or chemical substitution to switch magnetic order, enabling order-of-magnitude control over exciton lifetime and oscillator strength.

## Conclusion

In summary, cluster-assembled $V_4S_9X_4$ monolayers achieve the coexistence of above-room-temperature ferromagnetism and giant exciton binding, surpassing the conventional 2D magnetic semiconductors. GW-BSE calculations reveal an exciton binding energy of 1.85 eV in $V_4S_9Br_4$, which originates from ultra-flat bands, weak dielectric screening ($\varepsilon_\infty = 2.95$), and extreme real-space localization within individual clusters. While the ground state is a dark exciton, the first bright exciton emerges at higher energy, resulting in a pronounced lifetime gap between dark and bright states. Crucially, switching the magnetic order reversibly tunes excitons from a mixed Frenkel-Wannier state to a pure Frenkel state while preserving the giant binding energy. The darkest antiferromagnetic exciton achieves an ultra-long radiative lifetime of 23.95 μs, enabling wide-range magnetic modulation of exciton dynamics. This work establishes cluster assembly as a design paradigm for high-performance magneto-optoelectronics, promising for spin memory and quantum information technologies.

## Acknowledgements

This work was supported by the National Natural Science Foundation of China (12474271, 12534012 and 12574303) and Guangdong Provincial Quantum Science Strategic Initiative (GDZX2401002).

## Competing interests

The authors declare no competing interests.

Table 1. Lattice constant $a$ (Å), formation energy $E_f$ (eV/atom), ground-state magnetic configuration (GS), PBE and HSE06 band gaps (eV), magnetic moment M of each V atom ($\mu_B$), easy magnetization axis (EMA), magnetocrystalline anisotropy energy (MAE, μeV/atom), Heisenberg exchange parameters $J_1$–$J_4$ (meV), and Curie temperature $T_C$ (K) for $V_4S_9X_4$ (X = F, Cl, Br, I) monolayers.

| Material | $V_4S_9F_4$ | $V_4S_9Cl_4$ | $V_4S_9Br_4$ | $V_4S_9I_4$ |
|---|---|---|---|---|
| $a$ (Å) | 10.14 | 10.75 | 10.99 | 11.34 |
| $E_f$ (eV/atom) | –1.59 | –0.89 | –0.76 | –0.69 |
| GS | FM | FM | FM | FM |
| $E_g^{PBE}$ (eV) | 0.71 | 0.75 | 0.75 | 0.76 |
| $E_g^{HSE}$ (eV) | 1.06 | 1.22 | 1.27 | 1.41 |
| M ($\mu_B$) | 1.08 | 1.09 | 1.09 | 1.04 |
| EMA | 001 | 001 | 001 | 001 |
| MAE (μeV/atom) | 108.57 | 118.51 | 132.66 | 160.27 |
| $J_1$ (meV) | 22.84 | 15.06 | 13.16 | 10.33 |
| $J_2$ (meV) | 6.54 | 2.69 | 2.48 | 3.16 |
| $J_3$ (meV) | –0.75 | –4.17 | –2.69 | –0.59 |
| $J_4$ (meV) | 0.83 | 0.78 | 0.96 | 1.11 |
| $T_C$ (K) | 507.6 | 276.5 | 268.3 | 275.0 |

Table 2. Irreducible representations of the band-edge states at the Γ point for the ferromagnetic $V_4S_9Br_4$ monolayer, listed from VBM-3 to CBM+3 for both majority-spin (Spin-Up) and minority-spin (Spin-Down) channels. The degeneracy of each state is indicated. The VBM ($B_{1u}$) to CBM ($E_u$) transition is parity-forbidden in the majority-spin channel, while allowed transitions follow the $D_{4h}$ dipole selection rules.

| Irreducible representation | VBM-3 | VBM-2 | VBM-1 | VBM | CBM | CBM＋1 | CBM＋2 | CBM+3 |
|---|---|---|---|---|---|---|---|---|
| Spin-Up | $E_u$ | $A_{2u}$ | $B_{2g}$ | $B_{1u}$ | $E_u$ | $E_g$ | $A_{2u}$ | $A_{1u}$ |
| Degeneracy | 2 | 1 | 1 | 1 | 2 | 2 | 1 | 1 |
| Spin-Down | $E_u$ | $A_{1u}$ | $B_{2u}$ | $B_{1g}$ | $E_g$ | $A_{1g}$ | $E_u$ | $B_{2g}$ |
| Degeneracy | 2 | 1 | 1 | 1 | 2 | 1 | 2 | 1 |

Table 3. Radiative lifetimes of low-lying excitons in the FM and AFM configurations of the $V_4S_9Br_4$ monolayer. Dark excitons consistently exhibit lifetimes orders of magnitude longer than bright excitons, and AFM lifetimes are uniformly longer than their FM counterparts.

| FM | $E_S$ (eV) | $\mu_S^2$ (Bohr) | τ | AFM | $E_S$ (eV) | $\mu_S^2$ (Bohr) | τ |
|---|---|---|---|---|---|---|---|
| $D_I$ | 0.211 | $6.10\times10^{-3}$ | 1.20 ns | $D_I$ | 0.409 | $1.65\times10^{-3}$ | 2.29 ns |
| $D_{II}$ | 0.279 | $9.73\times10^{-3}$ | 0.57 ns | $D_{II}$ | 0.421 | $3.08\times10^{-3}$ | 1.19 ns |
| $B_I$ | 0.311 | 0.06 | 86.87 ps | $D_{III}$ | 0.426 | $3.49\times10^{-3}$ | 1.04 ns |
| $B_{II}$ | 0.334 | 0.01 | 0.40 ns | $D_{IV}$ | 0.443 | $3.87\times10^{-3}$ | 0.90 ns |
| $B_{III}$ | 0.483 | 0.11 | 28.24 ps | $D_V$ | 0.574 | $1.13\times10^{-7}$ | 23.95 μs |
| $B_{IV}$ | 0.494 | 0.01 | 0.29 ns | $D_{VI}$ | 0.580 | $3.79\times10^{-7}$ | 7.06 μs |
| $B_V$ | 0.565 | 0.03 | 82.61 ps | $D_{VII}$ | 0.595 | $5.87\times10^{-7}$ | 4.44 μs |
| $D_{III}$ | 0.593 | $1.13\times10^{-3}$ | 2.31 ns | $D_{VIII}$ | 0.596 | $3.64\times10^{-7}$ | 7.15 μs |
| $D_{IV}$ | 0.609 | $2.56\times10^{-4}$ | 9.95 ns | $D_{IX}$ | 0.636 | $1.98\times10^{-5}$ | 0.12 μs |
| $D_V$ | 0.628 | $1.02\times10^{-4}$ | 24.12 ns | $D_X$ | 0.638 | $1.11\times10^{-6}$ | 2.18 μs |
| $D_{VI}$ | 0.631 | $5.68\times10^{-6}$ | 0.43 μs | $B_I$ | 0.674 | 0.01 | 0.18 ns |

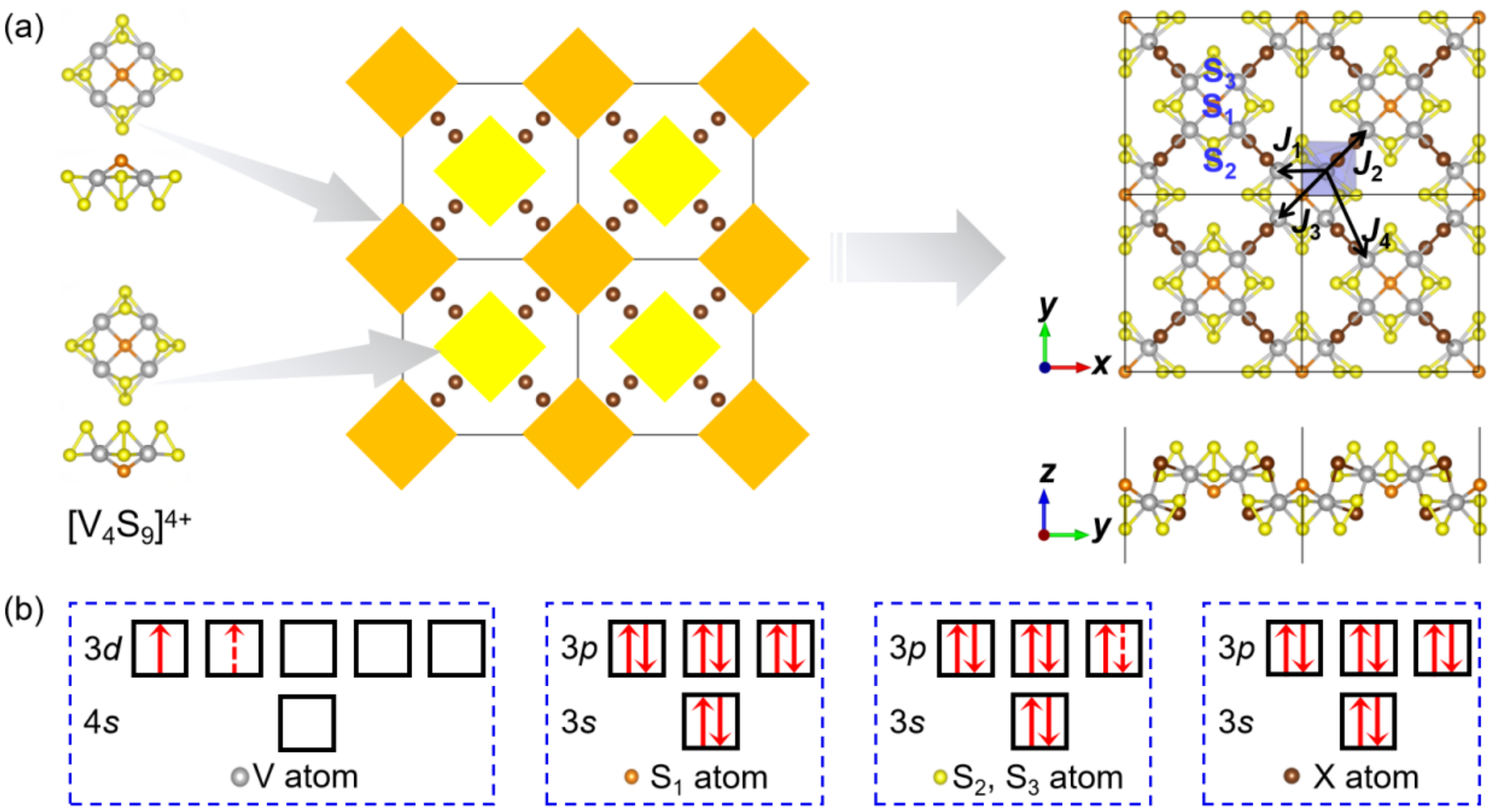

Fig. 1. (a) Schematic of the cluster-assembly strategy: discrete $[V_4S_9]^{4+}$ clusters linked by halide ligands $X^-$ (X=F, Cl, Br, I) into an extended $V_4S_9X_4$ monolayer. The four exchange pathways $J_1$–$J_4$ are indicated. (b) Valence electron filling diagram for each constituent atom.

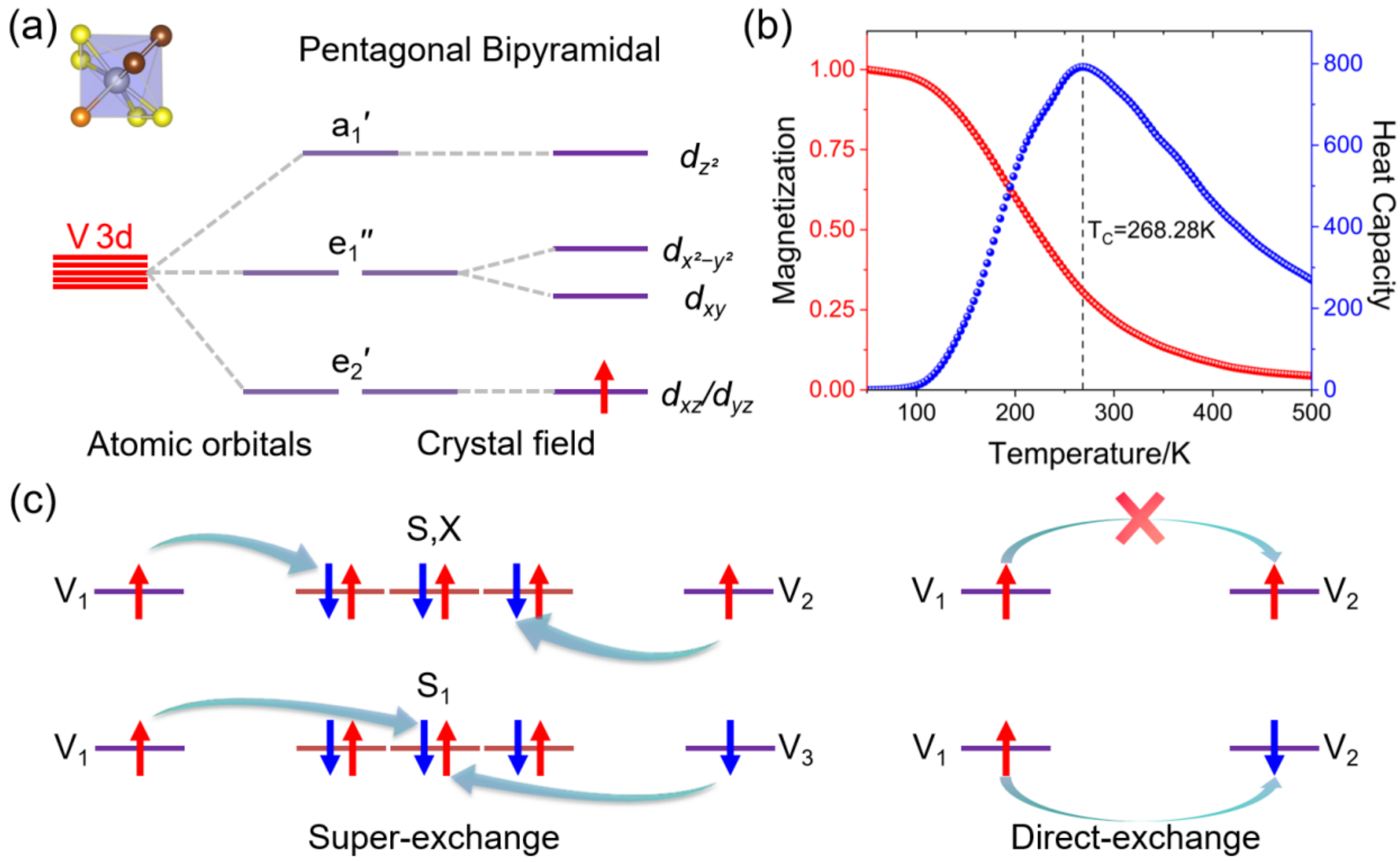

Fig. 2. (a) Crystal field splitting of V-3*d* orbitals under the $D_{5h}$ pentagonal bipyramidal coordination environment, showing the $a_1$' ($d_{z^2}$), $e_2$' ($d_{x^2-y^2}$,$d_{xy}$), and $e_1$'' ($d_{yz}$, $d_{xz}$) levels with the unpaired electron occupying the lowest-lying $e_1$'' orbital. (b) Heat capacity and magnetic moment as a function of temperature from Monte Carlo simulations for $V_4S_9Br_4$, yielding the Curie temperature $T_C$. (c) Schematic of the dominant superexchange pathways: $V_1$-S-$V_2$ and $V_1$-X-$V_2$ ($J_1$, ferromagnetic), $V_1$-$S_1$-$V_3$ ($J_3$, antiferromagnetic), and the direct V-V *d*-*d* exchange (negligible).

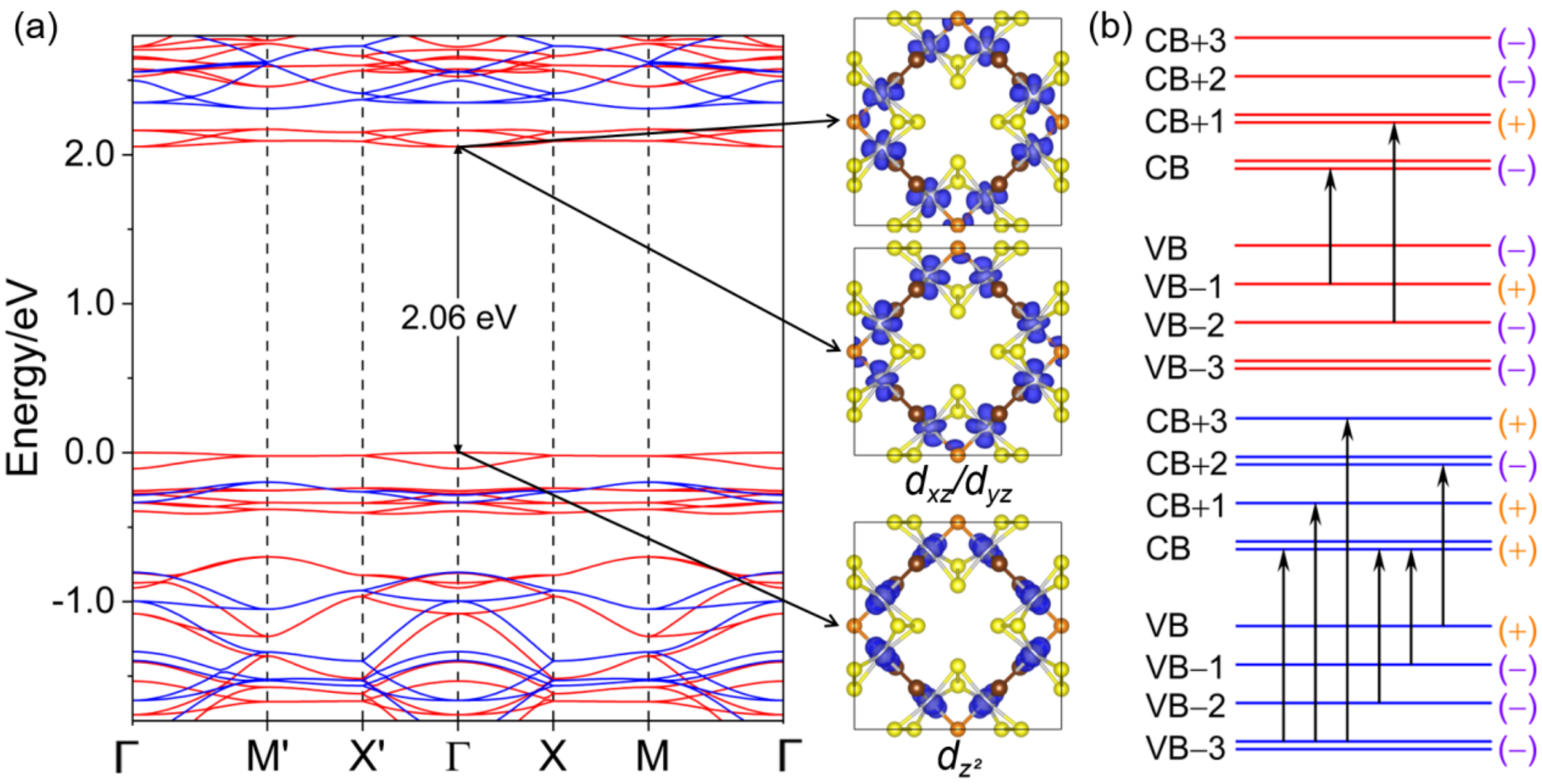

Fig. 3. (a) GW quasiparticle band structure of the ferromagnetic $V_4S_9Br_4$ monolayer along high-symmetry paths of the Brillouin zone. Majority-spin (spin-up, in red) and minority-spin (spin-down, in blue) bands are presented, with a direct majority-spin band gap of 2.06 eV located at the Γ point. Insets: charge density distributions of the VBM and CBM at Γ. (b) Irreducible representations of the band-edge states at Γ and the corresponding optical transition selection rules under $D_{4h}$ point group symmetry.

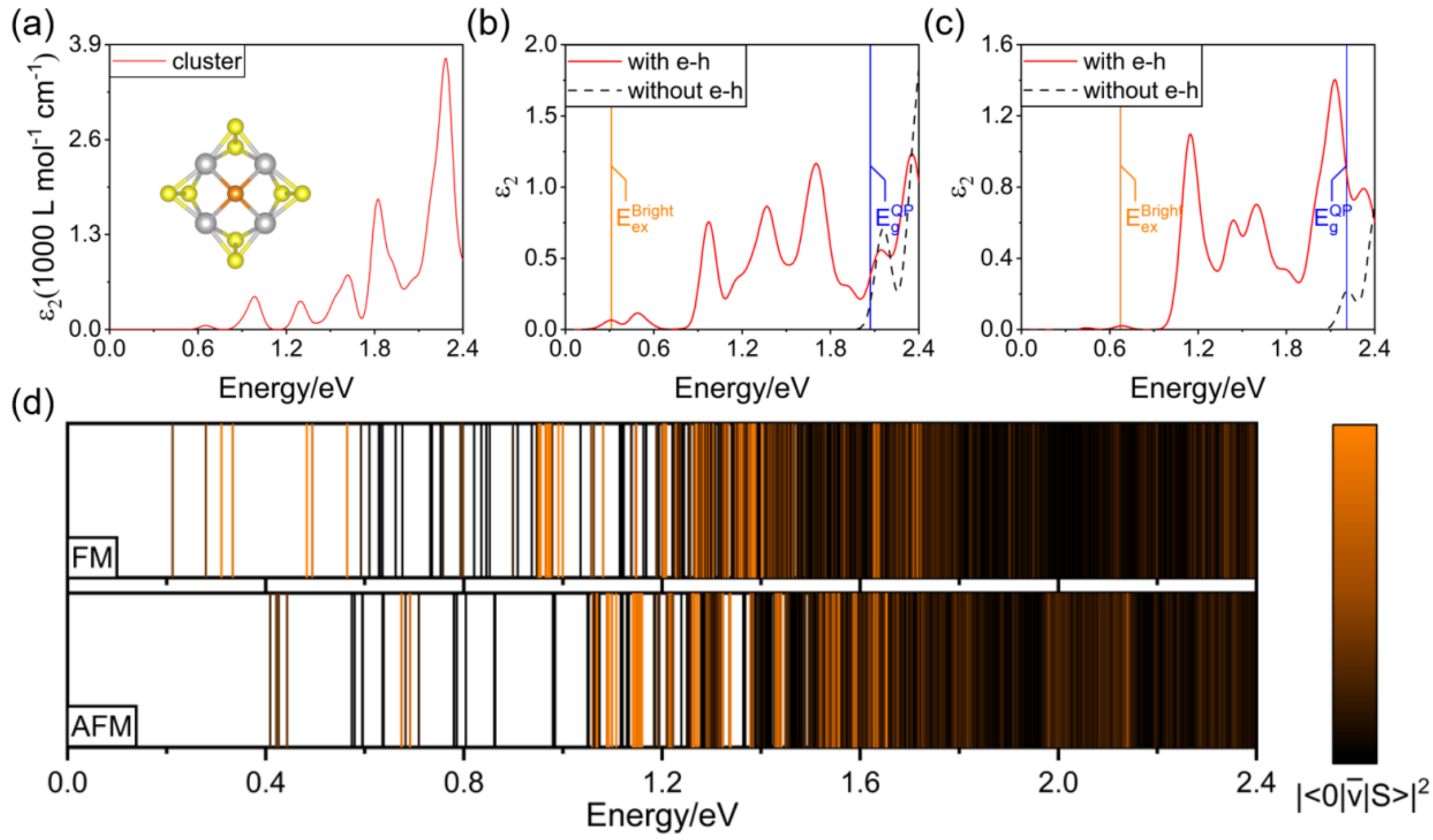

Fig. 4. (a) TD-DFT absorption spectrum of the isolated $[V_4S_9]^{4+}$ cluster. (b) Optical absorption spectra of the ferromagnetic $V_4S_9Br_4$ monolayer: independent particle approximation (IPA, dashed) and Bethe-Salpeter equation (BSE, solid), showing a ~1.1 eV redshift of the absorption edge. (c) Optical absorption spectra of the antiferromagnetic $V_4S_9Br_4$ monolayer (IPA vs. BSE). (d) Exciton eigenstate energies for the FM and AFM configurations. Transition dipole moment is indicated by the horizontal bar length. Dark excitons in black, bright excitons in orange.

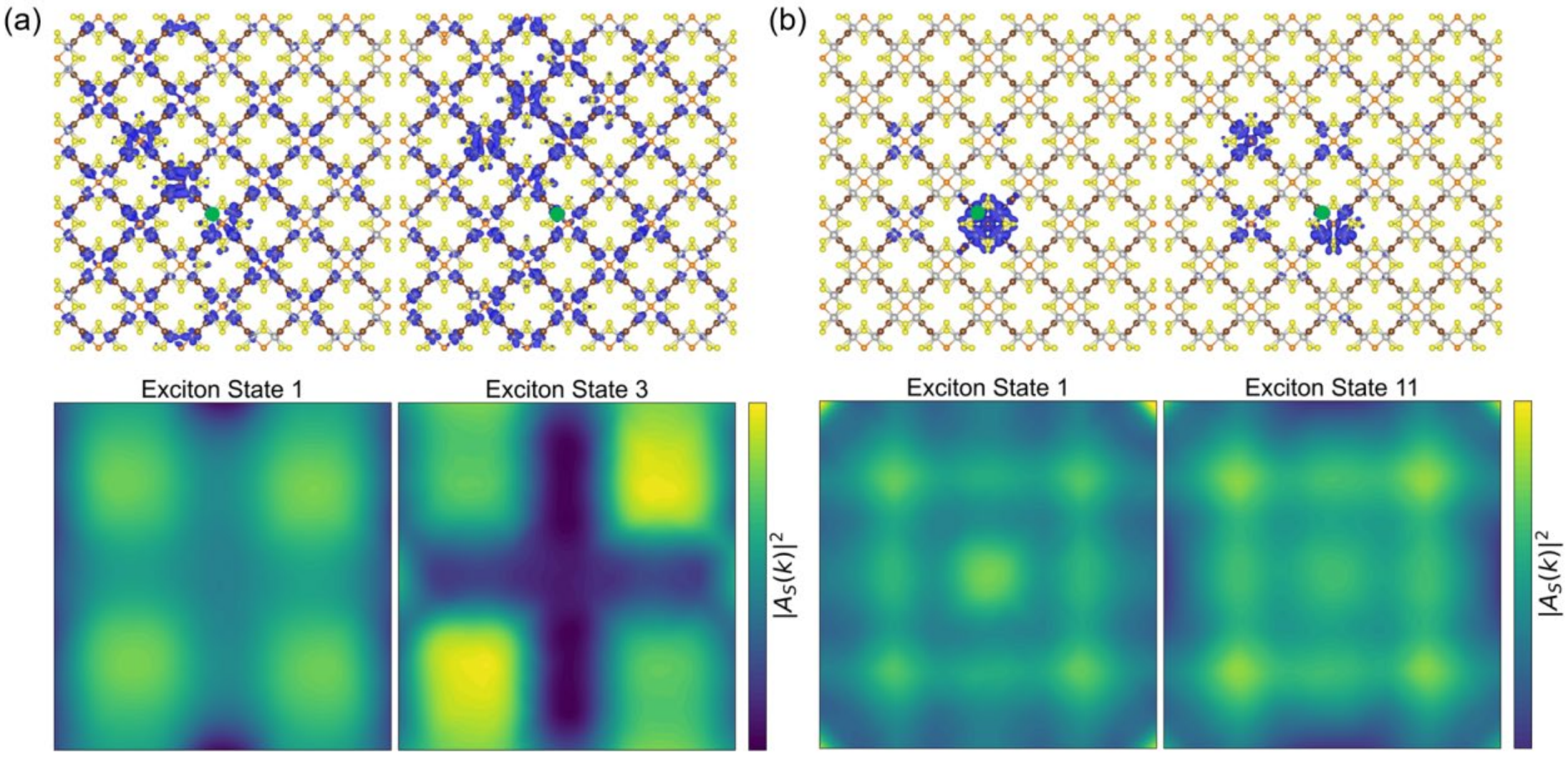

Fig. 5. (a) Ferromagnetic phase: real-space exciton density distribution (top) and reciprocal-space $|A_S(k)|^2$ distribution (bottom) for the dark exciton $D_I$ (State 1, left) and the bright exciton $B_I$ (State 3, right). The hole is fixed on the V atom marked in green. (b) Antiferromagnetic phase: reciprocal-space and real-space distributions for the lowest dark exciton (State 1) and the dark state at 0.57 eV (State 11). The color scale in (b) is compressed by a factor of six relative to (a) to render the weak residual modulation visible.

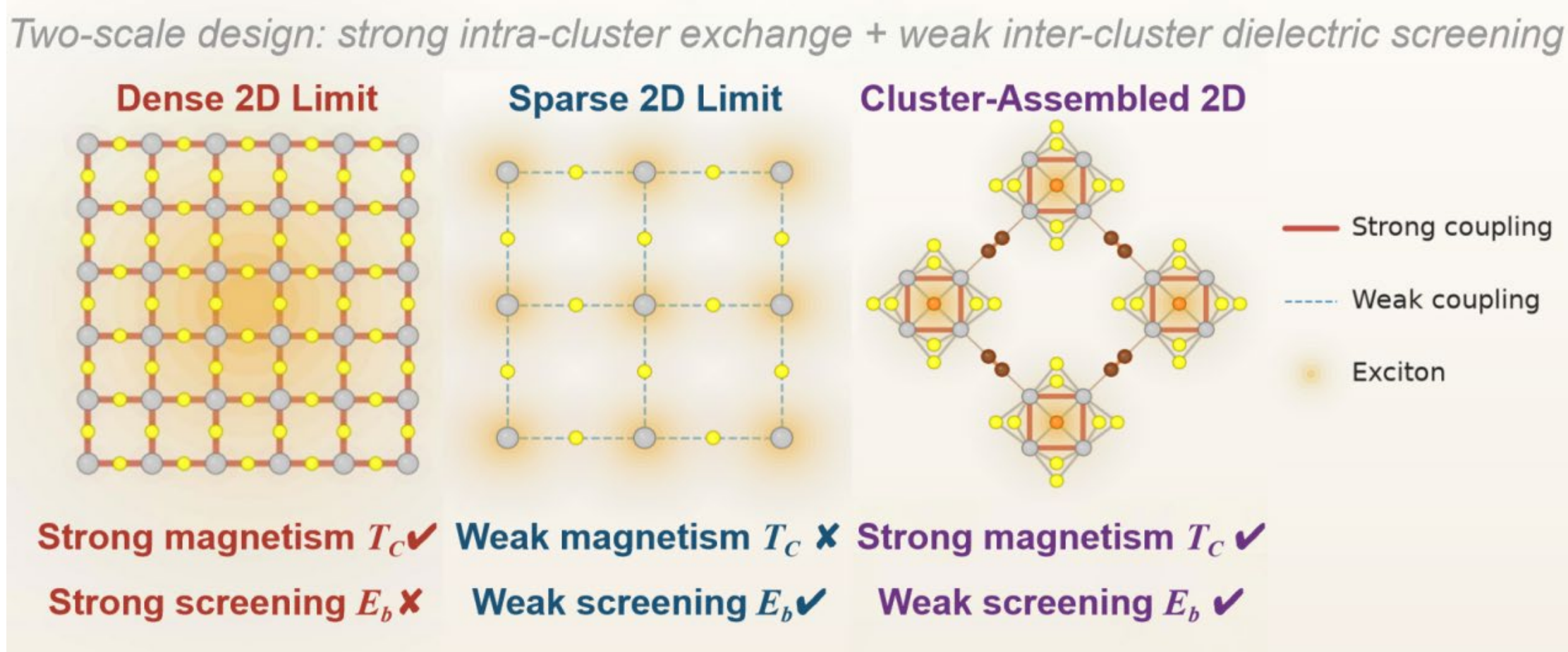

Fig. 6. Hierarchical architecture of cluster-assembled monolayers. Dense 2D systems (strong inter-site coupling) sustain long-range magnetic order, but their strong dielectric screening suppresses exciton binding. Sparse 2D systems (weak inter-site coupling) preserve strongly bound localized excitons, yet their insufficient coupling prevents collective magnetic order. Cluster assembly reconciles these conflicting trends via a hierarchical two-level structure: $[V_4S_9]^{4+}$ clusters host intra-cluster ferromagnetic superexchange and localized excitons, while $X^{-}$ bridges provide moderate inter-cluster coupling that mediates long-range magnetic order without quenching exciton binding through enhanced dielectric screening.

**Supplemental Materials**

# Giant exciton effects and magneto-excitonic coupling in $V_4S_9X_4$ 2D magnetic semiconductors

Yingjie Wei[1], Fan Zhang[2], Ying Zhao[3], Lixin Zhou[1], Yan Su[1], Yu Guo[1*], Jijun Zhao[4,5*]

[1] Key Laboratory of Materials Modification by Laser, Ion and Electron Beams (Dalian University of Technology), Ministry of Education, Dalian 116024, China

[2] School of Physics, Nanjing University, Nanjing 210093, China

[3] School of Electronic Engineering, Huainan Normal University, 232038, China

[4] Guangdong Provincial Key Laboratory of Quantum Engineering and Quantum Materials, School of Physics, South China Normal University, Guangzhou 510006, China

[5] Guangdong Basic Research Center of Excellence for Structure and Fundamental Interactions of Matter, South China Normal University, Guangzhou 510006, China

[*]Corresponding authors. Email: guoyu_dlut@dlut.edu.cn, zhaojj@scnu.edu.cn

## S1. Effect of spin-orbit coupling on the band structure

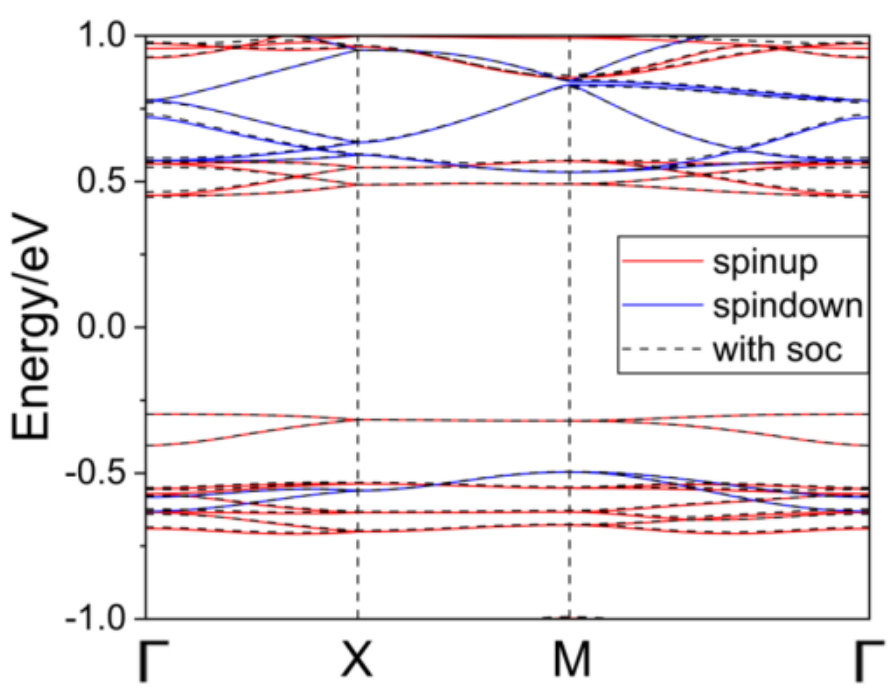

Fig. S1. GW band structure of ferromagnetic $V_4S_9Br_4$ comparing the scalar-relativistic calculation (spin-up: solid red; spin-down: solid blue) with the fully relativistic calculation including spin–orbit coupling (SOC, black dashed). The near-perfect overlap justifies neglecting SOC in the exciton calculations.

**S2. Convergence tests for excitonic calculations**

Table S1. Convergence behavior of the calculated excitonic properties.

| Dielectric matrix cut-off (Ry) | $N_v$ | $N_c$ | Coarse *k*-grid | Fine *k*-grid | Lowest-exciton Energy (eV) | Second lowest-exciton Energy (eV) |
|---|---|---|---|---|---|---|
| 6 | 6 | 8 | 3×3×1 | 8×8×1 | 0.0989 | 0.1149 |
| | | | | 10×10×1 | 0.0982 | 0.1152 |
| | 7 | 8 | 3×3×1 | 10×10×1 | 0.0277 | 0.0739 |
| 12 | 6 | 7 | 3×3×1 | 10×10×1 | 0.1705 | 0.1808 |
| | 6 | 8 | 3×3×1 | 10×10×1 | 0.1246 | 0.1434 |
| | 6 | 9 | 3×3×1 | 10×10×1 | 0.0978 | 0.1234 |
| | 7 | 8 | 3×3×1 | 10×10×1 | 0.0548 | 0.1025 |
| | 7 | 9 | 3×3×1 | 10×10×1 | 0.0194 | 0.0792 |
| | | | 4×4×1 | 8×8×1 | 0.2464 | 0.3104 |
| | | | | 10×10×1 | 0.2049 | 0.2725 |
| | | | | 12×12×1 | 0.2660 | 0.3302 |
| | | | 5×5×1 | 10×10×1 | 0.1833 | 0.2366 |
| | | | | 12×12×1 | 0.1702 | 0.2149 |
| 15 | 7 | 8 | 3×3×1 | 10×10×1 | 0.0781 | 0.1254 |
| | | | 4×4×1 | 10×10×1 | 0.2375 | 0.2988 |
| | 7 | 9 | 3×3×1 | 10×10×1 | 0.0428 | 0.1023 |
| | | | 4×4×1 | 10×10×1 | 0.2119 | 0.2792 |
| | | | | 12×12×1 | 0.2730 | 0.3367 |
| 18 | 7 | 9 | 4×4×1 | 8×8×1 | 0.2536 | 0.3173 |
| | | | | 10×10×1 | 0.2316 | 0.2989 |
| | | | | 12×12×1 | 0.2927 | 0.3564 |

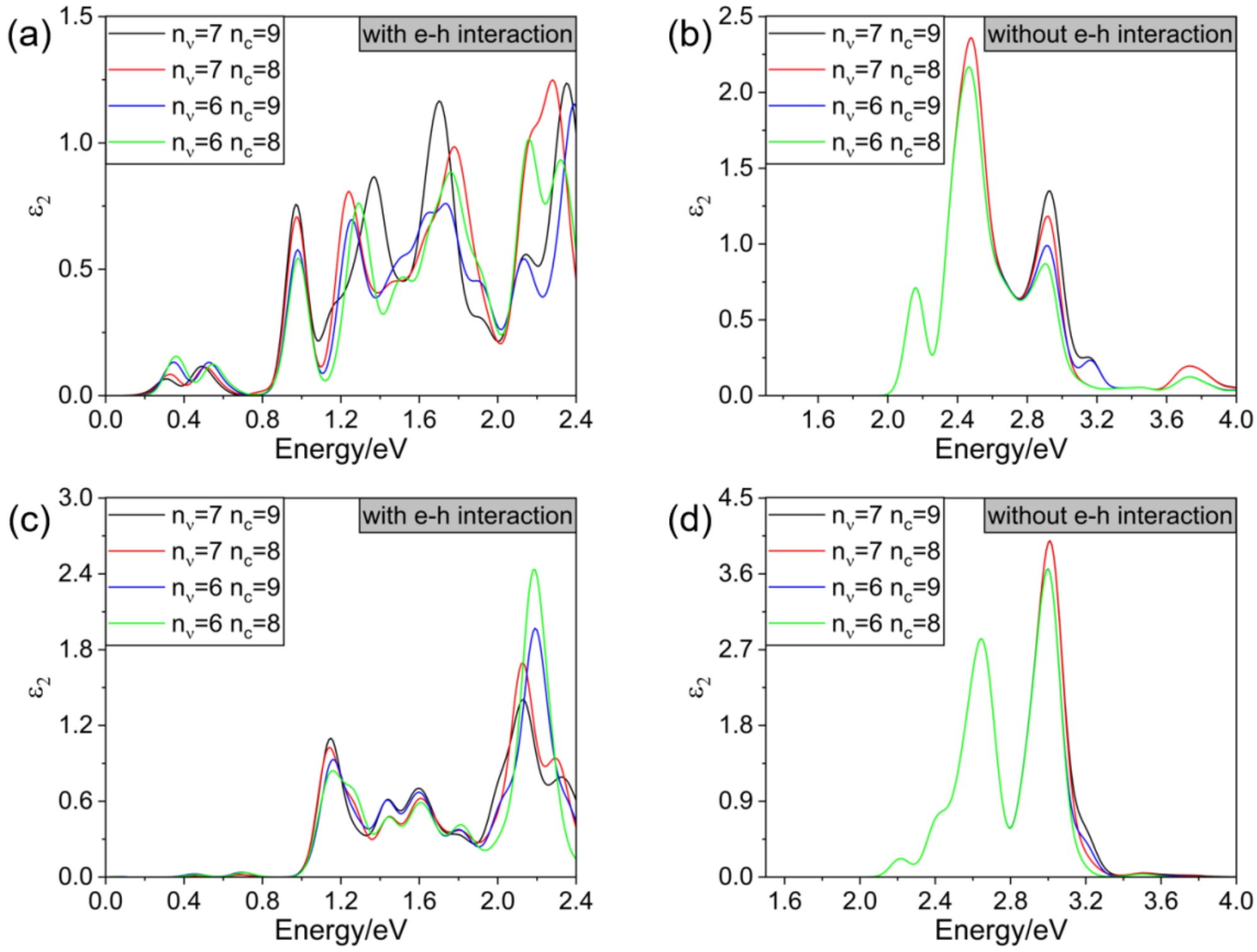

Fig. S2. Convergence tests of the exciton binding energy with respect to the number of valence and conduction bands included in the BSE kernel. (a) FM with electron-hole interaction (BSE), (b) FM without e-h interaction (IPA), (c) AFM with electron-hole interaction (BSE), (d) AFM without e-h interaction (IPA). The converged band numbers used in the production calculations are marked by arrows.

## S3. Stability of $V_4S_9X_4$ monolayers

To verify the experimental feasibility of these isostructural $V_4S_9X_4$ monolayers, we systematically evaluate their intrinsic stability from four dimensions: energetic, dynamic, thermal and mechanical exfoliation feasibility. First, the formation energy ($E_f$) is defined as follows:

$$E_f = (E_{tot} - n_1 \times E_V - n_2 \times E_S - n_3 \times E_X)/N \quad (1)$$

where $E_{tot}$ is the total energy of the monolayer per unit cell, $E_V$ and $E_S$ represent the energies per atom of bcc vanadium metal and orthorhombic α-sulfur, respectively; $E_X$ denotes the energy per atom of the corresponding isolated halogen diatomic molecule ($X_2$). The variables $n_1$, $n_2$ and $n_3$ denote the numbers of V, S and halogen atoms in the unit cell, respectively, and $N = n_1 + n_2 + n_3$ is the total number of atoms within the unit cell. The calculated $E_f$ values range from −1.59 to −0.69 eV/atom. All these exergonic (negative) formation energies confirm that the $V_4S_9X_4$ monolayers are thermodynamically stable relative to their elemental precursors, suggesting their synthetic accessibility under appropriate experimental conditions. Second, the phonon dispersion spectra for all monolayers exhibit no imaginary frequencies throughout the entire Brillouin zone (Fig. S3), corroborating their dynamic stability against spontaneous structural distortion. Third, *ab initio* molecular dynamics simulations performed at 300 K and 500 K over a 10 ps timescale (Fig. S4) reveal no bond breaking, atomic rearrangement, or structural collapse, thereby verifying their robust thermal stability at and well above room temperature. Fourth, to assess the feasibility of mechanically exfoliating monolayers from the experimentally available layered bulk $V_4S_9Br_4$ (the atomic structure exhibited in Fig. S5), we calculate the interlayer cohesive energy $E_c$, which is defined as the average atomic energy required to separate a monolayer from its bulk counterpart, as formulated below:

$$E_c = (E_{bulk} - E_{monolayer})/N \quad (2)$$

In this equation, $E_{bulk\ and}\, E_{monolayer}$ are the total energies of the bulk and monolayer unit cells, respectively, and $N$ is 34, the number of atoms in one monolayer unit cell. Three orthogonal views of the optimized bulk structure (Fig. S5) confirm well-defined van der Waals gaps between adjacent layers. The calculated $E_c$ of $V_4S_9Br_4$ monolayer is 54.2 meV/atom, which is comparable to the experimental interlayer cohesive energy of graphite (−52±5 meV/atom) and the calculated values of hexagonal boron nitride (56–60 meV/atom).[1] This low interlayer binding energy, together with the negative formation energies, absence of imaginary phonon modes, and robust thermal stability demonstrated above, confirms that the $V_4S_9Br_4$ monolayer is experimentally accessible via mechanical cleavage. Given the identical topological framework of the isostructural $V_4S_9X_4$ (X = F, Cl and I) series, similarly comparable interlayer binding strengths and exfoliation feasibility can be expected for the other compositions.

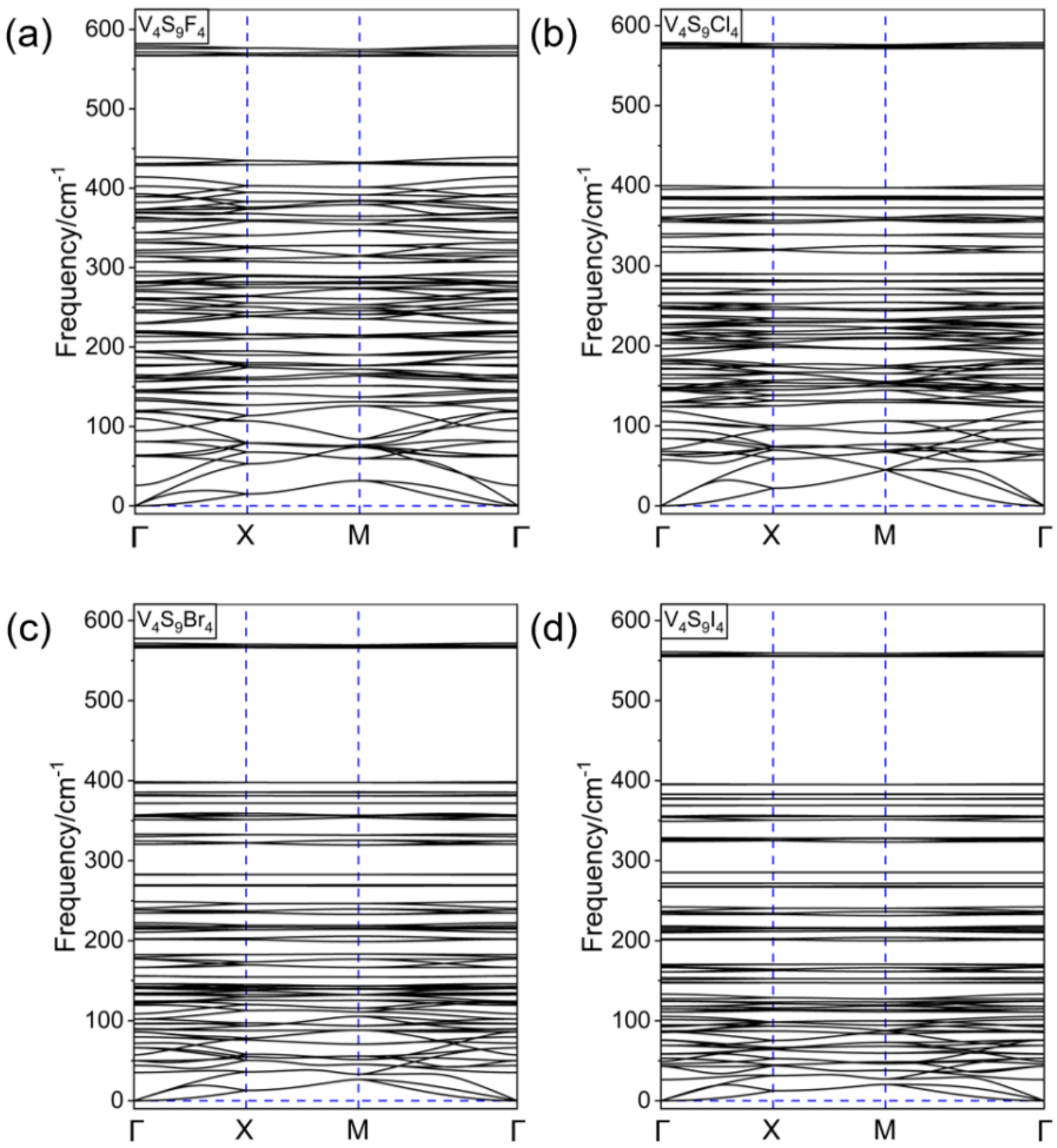

Fig. S3. Phonon dispersions for (a) $V_4S_9F_4$, (b) $V_4S_9Cl_4$, (c) $V_4S_9Br_4$, and (d) $V_4S_9I_4$ monolayers. The absence of imaginary frequencies in all four panels confirms the dynamical stability of the entire $V_4S_9X_4$ family.

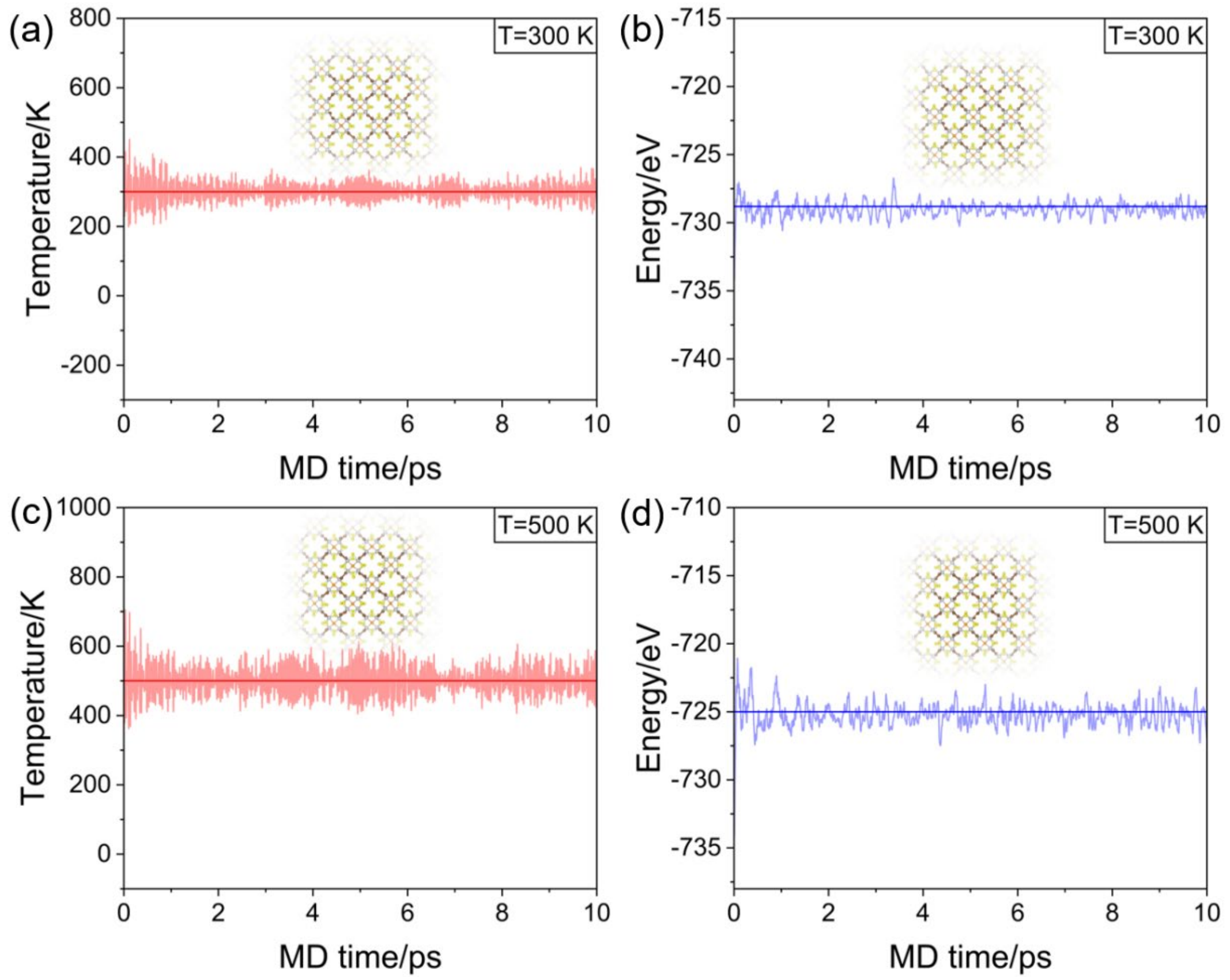

Fig. S4. *ab initio* molecular dynamics (AIMD) simulations of $V_4S_9Br_4$ monolayers at 300 K and 500 K over 10 ps. No bond breaking or structural collapse is observed, verifying thermal stability at and above room temperature.

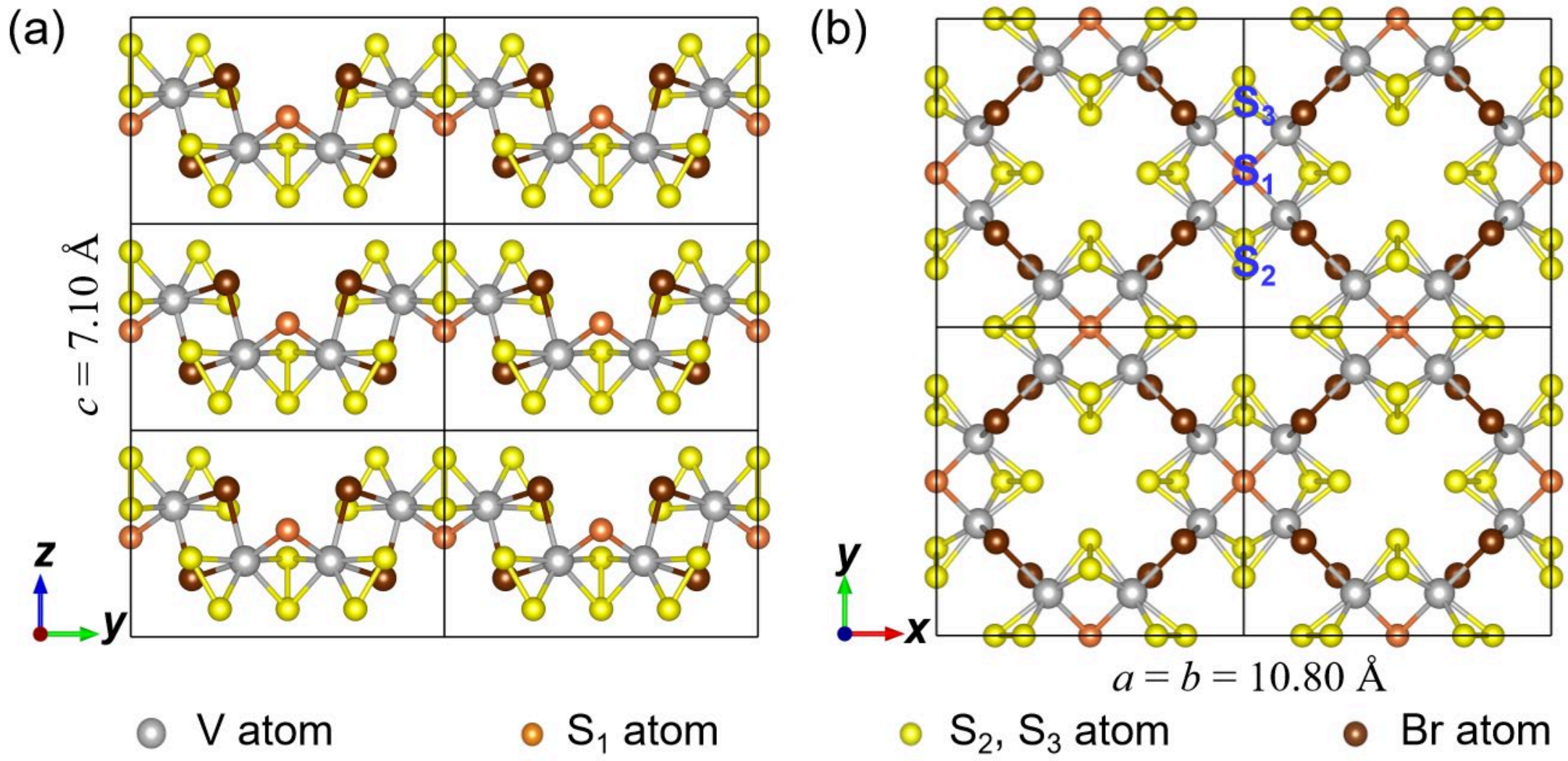

Fig. S5. Crystal structure of layered bulk $V_4S_9Br_4$ (space group P4/nmm). (a) Side view in the *bc* plane, illustrating the periodic stacking of monolayers with well-defined van der Waals gaps along the crystallographic *c*-axis. (b) Top view in the *ab* plane, showing square $[V_4S_9]^{4+}$ cluster units interconnected by bridging Br ligands. The DFT-optimized lattice parameters ($a$ = $b$ =10.80 Å, $c$ = 7.10 Å) are in good agreement with experimental single-crystal XRD data collected at 173 K ($a$ = $b$ = 10.86 Å, $c$ = 6.97 Å).[2]

**S4. Geometric parameters of magnetic exchange pathways in $V_4S_9X_4$ monolayers**

Table S2. V-S, V-X and V-V bond lengths $l$ (Å) and V-S-V and V-X-V bond angles $\theta$ (°) for $V_4S_9X_4$ monolayers, corresponding to the exchange pathways illustrated in Fig. 2c.

| Material | $V_1$-$S_1$-$V_2$ | | $V_1$-$S_2$-$V_2$ | | $V_1$-$S_3$-$V_2$ | | $V_1$-X-$V_2$ | | | V-V |
|---|---|---|---|---|---|---|---|---|---|---|
| | $\theta$ | $l$ | $\theta$ | $l$ | $\theta$ | $l$ | $\theta$ | $l$ | $\theta$ | $l$ |
| $V_4S_9F_4$ | 78.45 | 2.38 | 77.69 | 2.40 | 78.31 | 2.38 | 107.93 | 2.05 | 2.06 | 3.01 |
| $V_4S_9Cl_4$ | 77.98 | 2.37 | 77.19 | 2.39 | 77.91 | 2.37 | 99.30 | 2.47 | 2.49 | 2.98 |
| $V_4S_9Br_4$ | 77.87 | 2.36 | 76.95 | 2.39 | 77.53 | 2.37 | 97.76 | 2.49 | 2.63 | 2.97 |
| $V_4S_9I_4$ | 77.82° | 2.36 | 76.66 | 2.39 | 77.26 | 2.37 | 96.51 | 2.83 | 2.87 | 2.96 |

**S5. Electronic structures of $V_4S_9X_4$ monolayers**

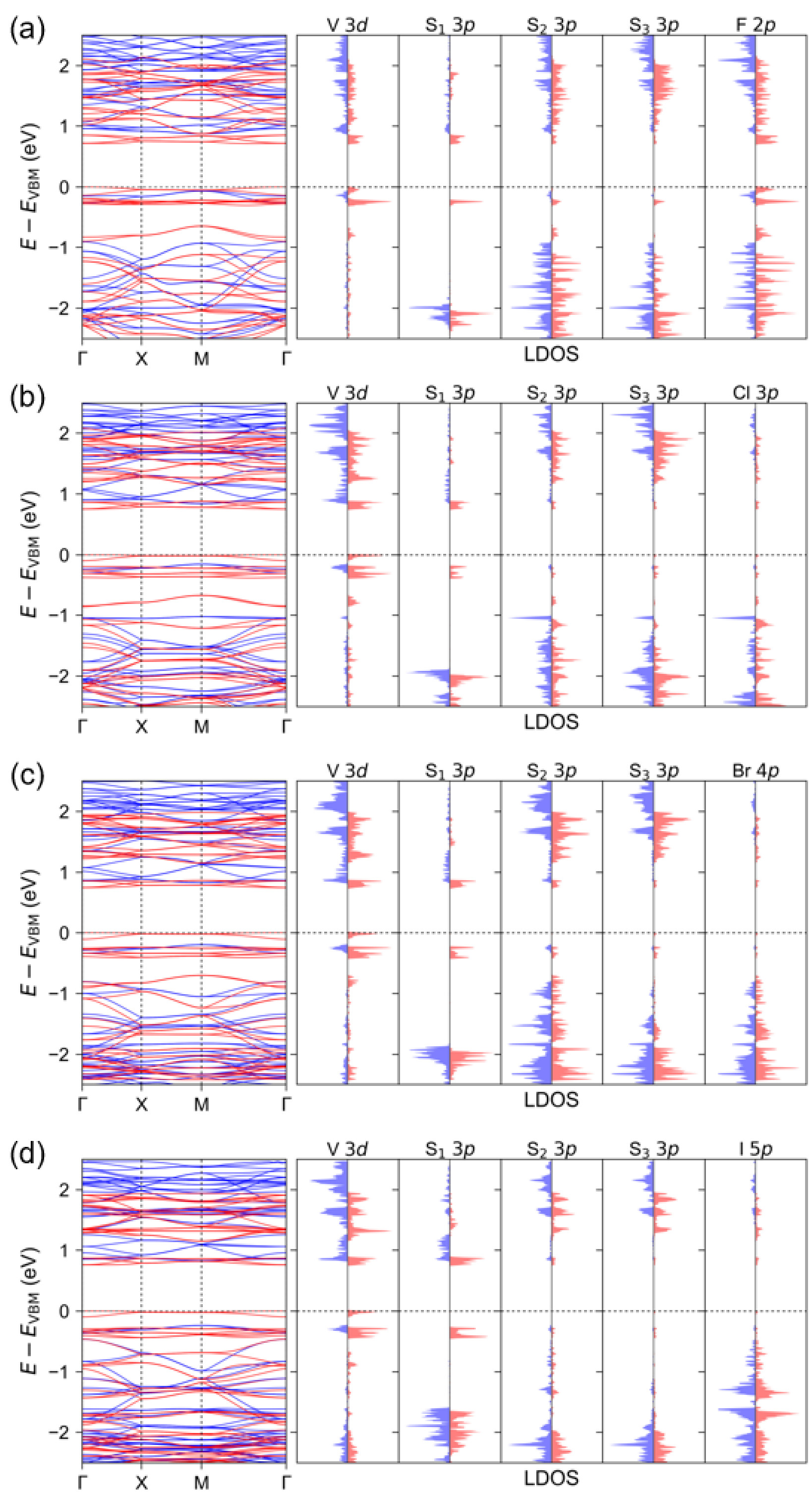

Fig. S6. PBE band structures and layer-resolved density of states (LDOS) for (a) $V_4S_9F_4$, (b) $V_4S_9Cl_4$, (c) $V_4S_9Br_4$, and (d) $V_4S_9I_4$ monolayers. The flat-band character and the dominant V-3d contribution to the band edges are preserved across the halogen series.

**S6. Magnetic configurations and relative stabilities of $V_4S_9X_4$ monolayers**

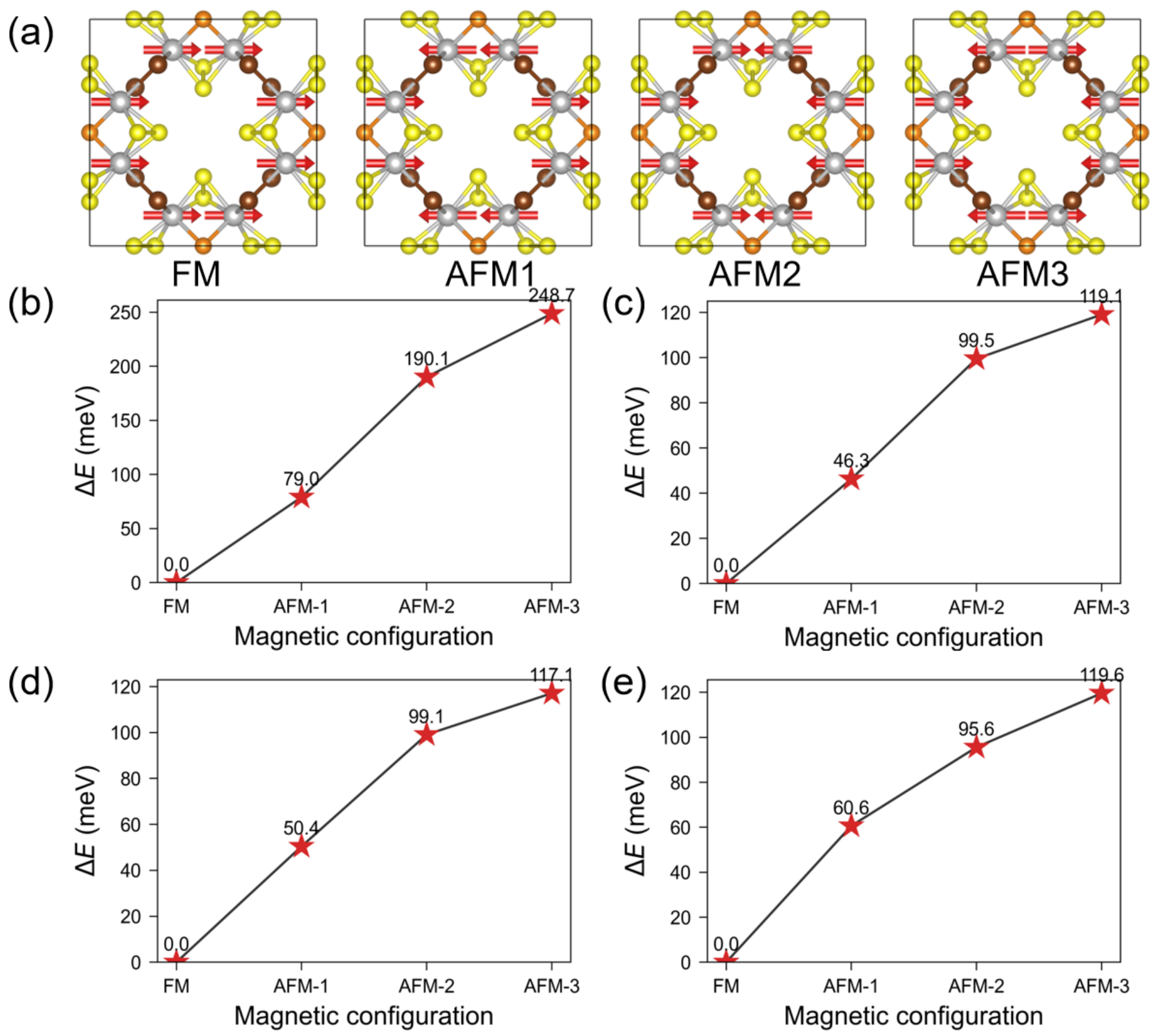

Fig. S7. (a) Schematic illustration of the ferromagnetic (FM) ground state and three symmetry-distinct antiferromagnetic (AFM) spin configurations considered in this work. (b)–(e) Total energy differences between the FM and AFM configurations for $V_4S_9F_4$, $V_4S_9Cl_4$, $V_4S_9Br_4$, and $V_4S_9I_4$, respectively. The FM configuration is the ground state for all four compositions.

**S7. Correlations between magnetic exchange coupling and structural parameters in $V_4S_9X_4$ monolayers**

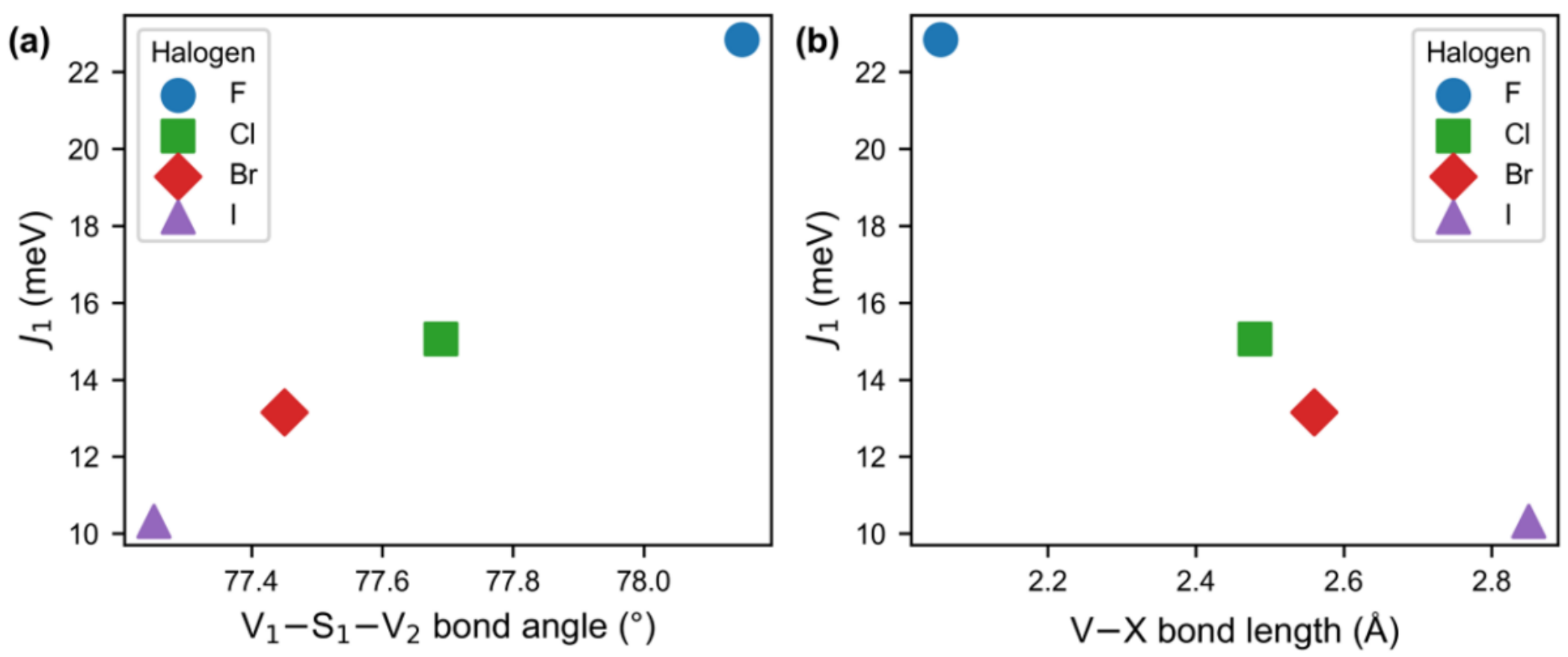

Fig. S8. (a) Correlation between the $V_1$-$S_1$-$V_2$ bond angle and the nearest-neighbor exchange coupling $J_1$ across the $V_4S_9X_4$ (X = F, Cl, Br, I) series, showing an approximately positive linear trend. (b) Correlation between the V-X bond length and $J_1$, showing an approximately negative linear trend, consistent with the dominant role of p-d orbital overlap in the superexchange mechanism.

## S8. Curie temperature of $V_4S_9X_4$ monolayers

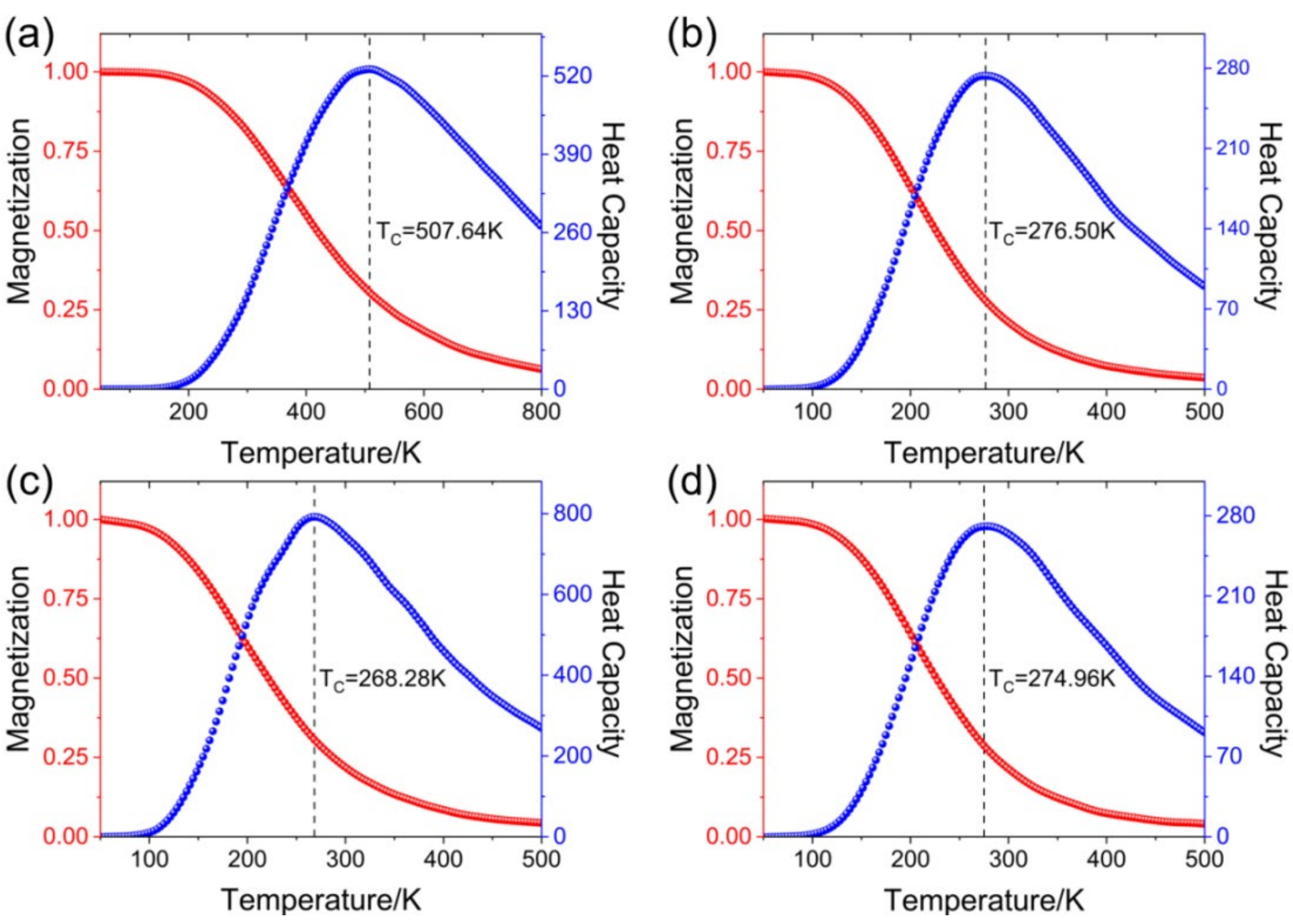

Fig. S9. Monte Carlo simulated magnetic susceptibility (left axis) and heat capacity (right axis) as a function of temperature for (a) $V_4S_9F_4$, (b) $V_4S_9Cl_4$, (c) $V_4S_9Br_4$, and (d) $V_4S_9I_4$ monolayers. The peak of the heat capacity defines the Curie temperature $T_C$ reported in Table 1.

**S9. Symmetry analysis of optical transitions**

Given the negligible SOC effect, spin remains a good quantum number, rendering the double-group formalism typically required for strongly spin–orbit-coupled systems unnecessary. We therefore employ the single-group approximation, in which spin-up and spin-down channels are treated separately within the same symmetry framework. This approach enables us to assign irreducible representations to each band-edge state purely on the basis of orbital symmetry, without mixing spin and spatial degrees of freedom. Within this scheme, we extract the symmetry characters of the four top valence bands and four bottom conduction bands at the Γ point. The corresponding irreducible representations (irreps) are assigned according to their dominant orbital contributions under single-group symmetry, as summarized in Table 2. For each spin channel (↑ and ↓), we evaluate the direct product of the irreps of the conduction and valence band states. An optical transition is allowed if the resulting direct product contains the irrep corresponding to the dipole operator for either in-plane or out-of-plane polarization.

For spin-up, the direct integral solutions of the irreducible representations of the unitary group allowed by the dipole are as follows:

$$B_{2g} \otimes E_u = E_u$$

$$A_{2u} \otimes E_g = E_u$$

For spin-down, the direct integral solutions of the irreducible representations of the unitary group allowed by the dipole are as follows:

$$E_u \otimes E_g = A_{1u} \oplus A_{2u} \oplus B_{1u} \oplus B_{2u}$$

$$E_u \otimes A_{1g} = E_u$$

$$E_u \otimes B_{2g} = E_u$$

$$A_{1u} \otimes E_g = E_u$$

$$B_{2u} \otimes E_g = E_u$$

$$B_{1g} \otimes E_u = E_u$$

All listed transitions are dipole-allowed, with some permitting both in-plane and out-of-plane polarization.

**S10. Electronic dielectric constants of several 2D materials**

Table S3. Comparison of the electronic dielectric constants $\varepsilon_\infty$ of $V_4S_9Br_4$ single layers, typical semiconductors Si and GaAs,[3] other two-dimensional exciton materials including $MoS_2$, $WS_2$, and h-BN,[4] and two molecular solids (Naphthalene and Anthracene).[5,6]

| | $V_4S_9Br_4$ | Si | GaAS | $MoS_2$ | $WS_2$ | *h*-BN | Naphthalene ($C_{10}H_8$) | Anthracence ($C_{14}H_{10}$) |
|---|---|---|---|---|---|---|---|---|
| $\varepsilon_\infty$ | 2.95 | 12.0 | 10.9 | 15.4 | 14 | 4.97 | 3.2 | 3.2 |

**S11. Exciton radiative lifetimes in FM and AFM $V_4S_9Br_4$ monolayer**

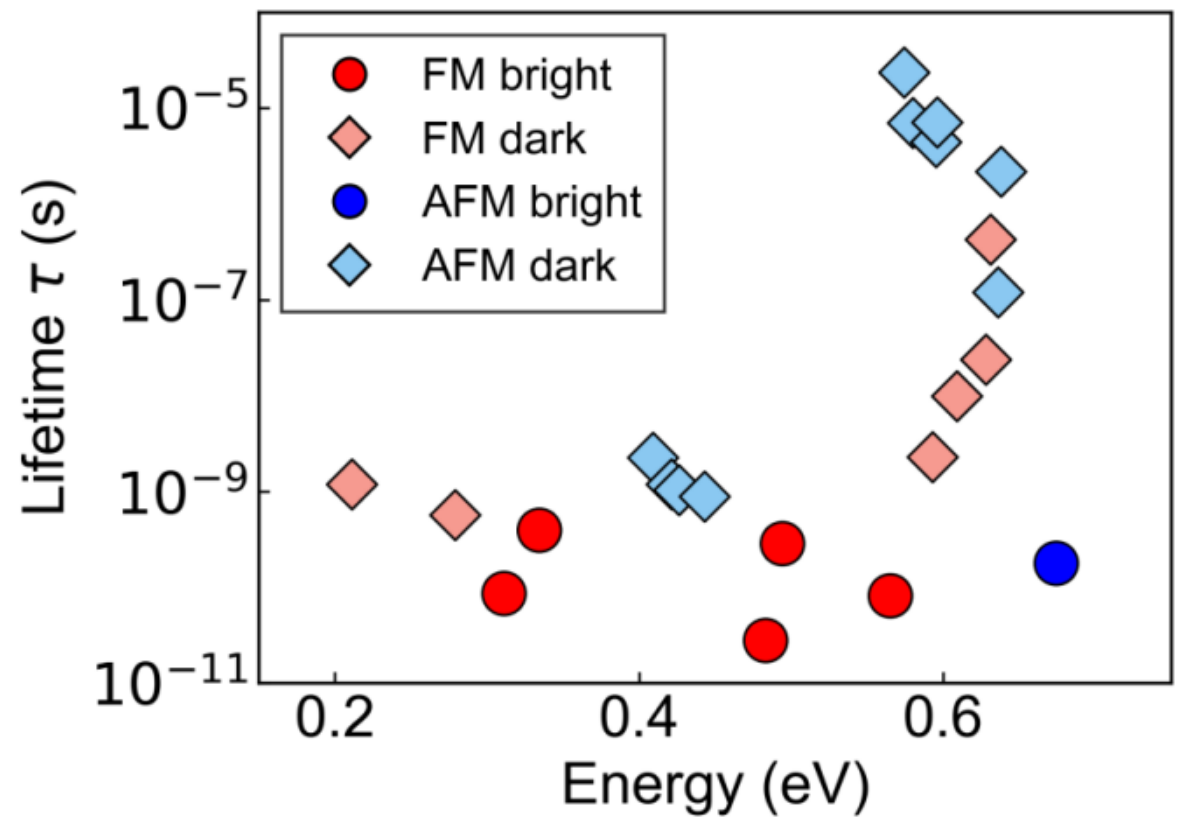

Fig. S10. Radiative lifetimes of low-lying exciton states in the FM (red) and AFM (blue) configurations of $V_4S_9Br_4$ monolayer, plotted against exciton energy on a logarithmic scale. Bright excitons (circles) and dark excitons (diamonds) are distinguished by their transition dipole moment. Dark excitons consistently exhibit lifetimes several orders of magnitude longer than bright excitons, and AFM lifetimes are uniformly longer than their FM counterparts at comparable energies.